\documentclass[english,prb,twocolumn,aps,amssymb,amsmath,superscriptaddress,showpacs]{revtex4-1}

\allowdisplaybreaks
\usepackage{array}
\usepackage{float}
\usepackage{graphicx,epsfig}
\usepackage[usenames]{color}
\usepackage{enumitem}   

\usepackage{hyperref}
\hypersetup{colorlinks=true}
\usepackage[all]{hypcap} 

\usepackage{csquotes}

\makeatletter

\providecommand{\tabularnewline}{\\}

\makeatother

\usepackage{babel}
\begin{document}

\title{Rotationally invariant slave-boson and density matrix
embedding theory: \\ A unified framework and a comparative study on the 1D and 2D Hubbard Model}


\author{Tsung-Han Lee}
\affiliation{Physics and Astronomy Department, Rutgers University, Piscataway,
New Jersey 08854, USA }
\author{Thomas Ayral}
\affiliation{Physics and Astronomy Department, Rutgers University, Piscataway,
New Jersey 08854, USA }
\affiliation{Atos Quantum Lab, Les Clayes-sous-Bois, France}
\author{Yong-Xin Yao}
\affiliation{Ames Laboratory-U.S. DOE and Department of Physics and Astronomy,
Iowa State University, Ames, Iowa 50011, USA}
\author{Nicola Lanata}
\affiliation{Department of Physics and Astronomy, Aarhus University, 8000,
Aarhus C, Denmark.}
\author{Gabriel Kotliar}
\affiliation{Physics and Astronomy Department, Rutgers University, Piscataway,
New Jersey 08854, USA }
\affiliation{Condensed Matter Physics and Materials Science Department,
Brookhaven National Laboratory, Upton, New York 11973, USA}

\begin{abstract}
We present detailed benchmark ground-state calculations of the one- and two-dimensional Hubbard
model utilizing the cluster extensions of the rotationally invariant slave-boson (RISB) mean-field
theory and the density matrix embedding theory (DMET). Our analysis shows that the overall
accuracy and the performance of these two methods are very similar. Furthermore, we
propose a unified computational framework that allows us to implement both of these techniques
on the same footing. This provides us with a new line of interpretation and paves the ways for
developing systematically new generalizations of these complementary approaches.

\end{abstract}
\maketitle

\section{Introduction}
Strongly  correlated electron systems are still  a most challenging problem in condensed-matter physics.  In  this area, quantum embedding approaches have proven to be invaluable tools for studying their electronic structure. In particular, dynamical mean-field theory (DMFT) \cite{Georges1996}, density matrix embedding theory (DMET) \cite{Knizia2012} and their respective cluster extensions have been successfully applied to many interacting model Hamiltonians as well as to real materials \cite{Georges1996,LDA_DMFT_RMP,Maier2005a,Hettler1998,Hettler1999,Lichtenstein2000,Kotliar2001,
Rohringer2017,Knizia2012,Knizia2013,Wouters2016a,Zheng2016,Zheng2017,Zheng_science_benchmark,
PRX_cluster_benchmark,PRX_H_chain_benchmark}.
The common basic idea underlying these schemes is to map the fully interacting lattice to a self-consistently determined impurity problem, for which a fragment of the original lattice, termed cluster, is treated as a correlated impurity coupled to a self-consistently determined non-interacting bath. 
%
The accuracy can be systematically improved by increasing the   reference cluster size towards the thermodynamic limit (TL) and the  size of the Hilbert space representing the non-interacting bath.

Another important theoretical method widely used for studying strongly correlated electron
systems is the rotationally-invariant slave-boson theory (RISB) \cite{Fresard1992,Lechermann2007,Lanata2016}, which 
is equivalent to the multi-orbital Gutzwiller approximation 
at the mean-field level \cite{Kotliar1986,Bunemann2007,Lanata_KLM} and 
generally provides predictions almost as accurate as DMFT~\cite{Isidori2009,Ferrero2008,Ferrero2009,Mazin2014,Lanata2015,Lanata2016,Piefke2017,Behrmann2015} (especially for the ground-state properties) while being much less computationally demanding.
Even if the foundation of the RISB mean-field theory is based on seemingly distinct ideas, it turns out that
also this framework can be viewed as a quantum-embedding theory. In fact, it has been recently shown~\cite{Lanata2015} that the RISB equations can be cast, similarly to DMET, in terms of ground-state calculations of auxiliary impurity systems named ``\emph{embedding Hamiltonians}",
whose non-interacting bath is determined self-consistently based on the
variational principle.
Subsequently, it has been also shown~\cite{Thomas_DMET_RISB} that DMET can be formally recovered from the RISB equation derived in Ref. \onlinecite{Lanata2016} 
by setting to unity the variational parameters encoding the mass renormalization weights. 


RISB and DMET are especially essential for the situations where the computational cost of DMFT becomes prohibitively large due to the exponentially growing Hilbert space and/or the sign problem in quantum Monte Carlo. This usually happens for the 5$f$ systems, where the crystal-field effects, spin-orbit-coupling interaction and lattice relaxation have to be taken into account simultaneously, and for the large-scale cluster simulations of the Hubbard model. Many challenging problems, such as the equations of state of elemental actinides and the phase diagram of the high $T_c$ superconductors, rely on such approximations to gain a qualitative or even quantitative understanding \cite{PRX_cluster_benchmark,Zheng_science_benchmark,Lanata2015}. Hence, it is of important interest to characterize the respective accuracy and performance of these two approaches.
 
Here we perform comparative RISB and DMET benchmark calculations on the 1D and 2D Hubbard model against the available exact solution and the DMET values extrapolated to the TL.
\cite{Zheng2017,PRX_cluster_benchmark}
Our numerical results indicate that the accuracy and the performance of these two methods are very similar for all the quantities studied, \emph{e.g.}, the total energy and local observables. 
Small differences between the two methods are found only for small cluster sizes, where RISB provides slightly more accurate predictions for the local observables (such as occupancy, double occupancy and local moments) as well as for the metal-insulator transition in the 2D Hubbard model.

Finally, we derive an alternative numerical implementation of DMET featuring a modified
RISB algorithm with mass renormalization weights set to
unity \cite{Thomas_DMET_RISB}, which provides us with a new line of interpretation and paves the way for developing new generalizations and synergistic combination of these approaches (\emph{e.g.}, to systems at finite temperature and/or with inter-site electron-electron interactions or electron-phonon interactions \cite{Lanata_GA_finite_T,Wang_GA_finite_T,Sandri_finite_T,PRX_H_chain_benchmark,DMET_e_phonon,DMET_e-p_Reinhard}). This implementation makes it also possible to pattern an interface between density functional theory (DFT) and DMET after previous DFT+RISB and DFT+DMFT works \cite{Lanata2015,LDA_DMFT_RMP}.

The paper is organized as follows: The Hubbard model is introduced in Sec. II. The RISB and DMET formalism and algorithmic structure are outlined in Sec. III.  In Section IV are presented our benchmark simulation of the Hubbard model in 1D and 2D. Finally, Sec. V is devoted to concluding remarks.

\section{Model}

Let us consider the 1D and 2D Hubbard model with the nearest neighbor hopping,

\begin{equation}
H=t\sum_{\sigma,\langle i,j\rangle} c^\dagger_{i\sigma}c_{j\sigma} + \sum_{i} U n_{i\uparrow}n_{i\downarrow},
\end{equation}

\noindent where $t$ is the hopping amplitude, $i$ and $j$ are the indices for the lattice sites, and the $\sigma$ is the spin label, and $U$ is the local Coulomb interaction. $c_{i\sigma}^{(\dagger)}$ is the annihilation (creation) operator for the electron at site $i$ and spin $\sigma$.

The cluster extensions of RISB and DMET are both implemented by tiling the original lattice with clusters of
increasing size \cite{Maier2005a}.
Thus, the degrees
of freedom of the single-band Hubbard model belonging to each cluster are treated as a single impurity, \emph{i.e.},
as if they were elementary (orbital) degrees of freedom
of a multi-orbital Hubbard Hamiltonian represented
as follows:

\begin{equation}
H=\sum_{\langle ij\rangle,\alpha,\beta}\tilde{t}_{ij}^{\alpha\beta}c_{i\alpha}^{\dagger}c_{j\beta}+\sum_{i}H_{\text{loc}}[\{c_{i\alpha},c_{i\alpha}^{\dagger}\}],\label{eq:Hubbard_rspace}
\end{equation}

\noindent where the indices $i,\ j=1,...,\mathcal{N}/N_{c}$ denote
the enlarged unit cell, $\mathcal{N}$ is the total number of atoms and $N_c$ is the number of atoms within each cluster and the labels
$\alpha,\ \beta=1,...,2N_{c}$ indicate the cluster spin and atom degrees of freedom. 

In order to utilize the RISB and DMET theory, it is useful to define the inter-cluster hopping matrix as follows:

\begin{equation}
\tilde{t}_{ij}^{\alpha\beta}=\Big\{\begin{array}{c}
t_{ij}^{\alpha\beta}\ \ \ \text{if}\ i\neq j\\
\ \ \ \ 0\ \ \ \ \ \text{otherwise}
\end{array}.\label{eq:hopping}
\end{equation}

\noindent The terms corresponding to the intra-cluster hopping parameters $t_{i\alpha,i\beta}$ are included within the operator $H_{\text{loc}}[\{c_{i\alpha},c_{i\alpha}^{\dagger}\}]$, along with the chemical potential and the local Coulomb interaction.

In our calculations, the translational invariance is exploited only partially, \emph{i.e.}, we represent the hopping matrix defined as:


\begin{equation}
\tilde{\varepsilon}_{\mathbf{k}}^{\alpha\beta}=\underset{i}{\sum}e^{-i\mathbf{k}\cdot \mathbf{r}_{i}}\tilde{t}_{i0}^{\alpha\beta},\label{eq:cluster_FT}
\end{equation}

\noindent where the momentum $\mathbf{k}$ belongs to the reduced Brillouin zone (RBZ) of the enlarged unit cell containing the cluster. The resulting Hamiltonian in the momentum space is represented as follows:

\begin{equation}
H=\sum_{\mathbf{k}\in \text{RBZ},\alpha,\beta}\tilde{\varepsilon}_{\mathbf{k}}^{\alpha\beta}c_{\mathbf{k}\alpha}^{\dagger}c_{\mathbf{k}\beta}+\sum_{\tilde{i}}H_{\text{loc}}[\{c_{i\alpha},c_{i\alpha}^{\dagger}\}],\label{eq:Hubbard_kspace}
\end{equation}

\noindent where $H_{\text{loc}}[\{c_{i\alpha,}c_{i\alpha}^{\dagger}\}]$ contains all the local one- and two-body terms.


\section{Methods}


As shown in Refs. \onlinecite{Lanata2015,Thomas_DMET_RISB,Knizia2012}, the RISB and DMET ground-state solution of the Hubbard Hamiltonian [Eq. \eqref{eq:Hubbard_kspace}] is obtained by solving recursively two
auxiliary systems: (i) a non-interacting system termed
``\emph{effective-medium}" or ``\emph{quasiparticle Hamiltonian}" and (ii) an interacting embedding
impurity problem called ``\emph{embedding Hamiltonian}."

The structure of the effective-medium Hamiltonian is the following:

\begin{equation}
H_{eff} = \sum_{\mathbf{k}\in \text{RBZ}} \Big[ R_{a\alpha}\tilde{\varepsilon}_{\mathbf{k}}^{\alpha\beta}R^{\dagger}_{\beta b}+\lambda_{ab}\Big] f_{\mathbf{k} a}^\dagger f_{\mathbf{k} b}, 
\label{eq:Hqp}
\end{equation}

\noindent where $\tilde{\varepsilon}_{\mathbf{k}}$ was defined in Eq. \eqref{eq:cluster_FT}, $R$ and $\lambda$ are $2N_c \times 2N_c$ complex matrices (the factor 2 arises from the spin degrees of freedom) and $\lambda$ is Hermitian. 
As we are going to show in Sec. \ref{risb-algorithm}, in RISB both $R$ and $\lambda$ are determined self-consistently \cite{Lanata2016} and their converged entries are connected to the self-energy $\Sigma(\omega)$ as follows: \cite{Lechermann2007,ghost_GA}

\begin{equation}
\Sigma(\omega)=-\omega\frac{1-R^\dagger R}{R^\dagger R}+\frac{1}{R}\lambda\frac{1}{R^\dagger}.\label{SE_RISB}
\end{equation}

\noindent On the other hand, in DMET only the entries of $\lambda$ (called $u$ in the DMET literature) can vary while $R=\mathbf{1}$, \emph{i.e.}, the self-energy consist exclusively of the part representing the on-site energy shifts: \cite{Knizia2012}

\begin{equation}
\Sigma( \omega )=\lambda,\label{SE_DMET}
\end{equation}

\noindent see Sec. \ref{risb-algorithm}.

The embedding Hamiltonian describes a multi-orbital dimer molecule containing a correlated impurity $c_{\alpha}^{(\dagger)}$ and a non-correlated bath $f_{a}^{(\dagger)}$. It reads:

\begin{eqnarray}
&& H_{\text{emb}}= H_{\text{loc}}\big[\{c_{\alpha}^{\dagger},c_{\alpha}\}\big] 
 \nonumber\\
&&\qquad +\sum_{\alpha a}\big(\mathcal{D}_{a\alpha}c_{\alpha}^{\dagger}f_{a}+\text{H.c.}\big)+\sum_{ab}\lambda_{ab}^{c}f_{b}f_{a}^{\dagger},\label{eq:Hemb}
\end{eqnarray}

\noindent where $H_{\text{loc}}$ is defined in Eq. \eqref{eq:Hubbard_rspace}, $\mathcal{D}$ and $\lambda^c$ are $2N_c \times 2N_c$ complex matrices and $\lambda^c$ is Hermitian. The entries of both matrices are determined self-consistently \cite{Knizia2012,Lanata2015,Lanata2016,Thomas_DMET_RISB}, see Secs. \ref{risb-algorithm} and  \ref{dmet-algorithm}. After convergence, the reduced density matrix of the impurity degrees of freedom (which is formally obtained by tracing out the bath degrees of freedom) provides the local reduced density matrix of the original physical system. In other words, the expectation value of any local operator $\hat{O}\big[\{c_{\alpha}^\dagger,c_{\alpha}\}\big]$, such as the double occupancy or the local stagger magnetic moment, can be calculated from the ground state wavefunction $|\Phi\rangle$ of $H_{\text{emb}}$ as follows:\cite{Lanata2015}

\begin{equation}
\langle O \rangle=\langle\Phi| \hat{O}\big[\{c_{\alpha}^\dagger,c_{\alpha}\}\big] |\Phi\rangle\,. \label{eq:local_observables}
\end{equation}

\subsection{Rotationally invariant slave-boson mean-field theory}\label{risb-algorithm}

The RISB theory is, in principle, an exact reformulation of the Hubbard system constructed by introducing auxiliary ``\emph{slave}" bosons coupled to ``\emph{quasiparticle}" fermionic degrees of freedom.~\cite{Lechermann2007,Lanata2015,Lanata2016}
As shown in Ref.~\onlinecite{Lanata2015}, 
the RISB mean-field theory
is entirely encoded in the following Lagrange function:


\begin{widetext}
\begin{eqnarray}
&& \mathcal{L}[|\Phi\rangle,R,\lambda,\Delta^{p};E^{c},\mathcal{D},\lambda^{c}]
=\nonumber\\ 
&&-\frac{1}{\beta}\frac{N_c}{\mathcal{N}}\sum_{\mathbf{k}\in \text{RBZ}}\sum_{i\omega_n}\text{Tr log}\big[i\omega_n\mathbf{1}-R_{a\alpha}\tilde{\varepsilon}_{\mathbf{k}}^{\alpha\beta}R^{\dagger}_{\beta b}-\lambda_{ab}\big]e^{i\omega_n0^{+}} +\sum_{i}\text{Tr}\Big[E^{c}(\langle\Phi|\Phi\rangle-1)+\langle\Phi|H_{\text{emb}}|\Phi\rangle\Big]
\nonumber \\
&&-\sum_{iab}\big(\lambda_{ab}+\lambda_{ab}^{c}\big)\Delta_{ab}^{p} -\sum_{ica\alpha}\big(\mathcal{D}_{a\alpha}R_{c\alpha}+c.c\big)\big[\Delta^{p}(1-\Delta^{p})\big]_{ca}^{1/2},
\label{eq:emb_action}
\end{eqnarray}
\end{widetext}

\noindent where:  $R$ and $\lambda$ are the renormalization coefficients
of the quasiparticle Hamiltonian
introduced in Eq.~\eqref{eq:Hqp}, $H_{\text{emb}}$, $\mathcal{D}$ and $\lambda^c$ are the parameters
of the embedding Hamiltonian introduced in Eq.~\eqref{eq:Hemb}, $|\Phi\rangle$ is the ground state wavefunction of $H_{\text{emb}}$, $E^c$ is a Lagrange multiplier enforcing the normalization of $|\Phi\rangle$ and 
$\Delta^p$ is the local density matrix of $H_{\text{eff}}$ (see Eq. \eqref{eq:SB_SP_eq1}).

The self-consistency conditions determining the parameters of $H_{\text{emb}}$ and $H_{\text{eff}}$, see Eqs.~\eqref{eq:Hqp} and ~\eqref{eq:Hemb}, are obtained by extremizing the mean-field Lagrange function with respect to $|\Phi\rangle,\ R,\ \lambda,\ \Delta^{p},\ E_{c},\ \mathcal{D},\ \text{and\ }\lambda^{c}$,
which leads to the following equations:

\begin{widetext}
\begin{eqnarray}
&&\Delta_{ab}^{p}=\frac{N_c}{\mathcal{N}}\sum_{\mathbf{k}\in \text{RBZ}}\big[f_T(R\tilde{\varepsilon}_{\mathbf{k}}R^{\dagger}+\lambda)\big]_{ba},\label{eq:SB_SP_eq1}\\
&&\big[\Delta^{p}(1-\Delta^{p})\big]^{1/2}_{ac}\mathcal{D}_{c\alpha}=\frac{N_c}{\mathcal{N}}\sum_{\mathbf{k}\in \text{RBZ}}\big[\tilde{\varepsilon}_{\mathbf{k}}R^{\dagger}f_T(R\tilde{\varepsilon}_{\mathbf{k}}R^{\dagger}+\lambda)\big]_{\alpha a},\label{eq:SB_SP_eq2}\\
&&\sum_{cb\alpha}\frac{\partial}{\partial d^p_s}\big[\Delta^p(1-\Delta^p)\big]^{\frac{1}{2}}_{cb}\big[\mathcal{D}\big]_{b\alpha}\big[R\big]_{c\alpha}+\text{c.c}+\big[l+l^c\big]_s=0,\label{eq:SB_SP_eq3}\\
&&H_{\text{emb}}|\Phi\rangle=E^{c}|\Phi\rangle,\label{eq:SB_SP_eq4}\\
&&\Big[\mathcal{F}^{(1)}\Big]_{ab}\equiv\langle\Phi|f_{b}f_{a}^{\dagger}|\Phi\rangle-\Delta_{ab}^{p}=0,\label{eq:SB_SP_eq5}\\
&&\Big[\mathcal{F}^{(2)}\Big]_{\alpha a}\equiv\langle\Phi|c_{\alpha}^{\dagger}f_{a}|\Phi\rangle-R_{c\alpha}\big[\Delta^{p}(1-\Delta^{p})\big]_{ca}^{\frac{1}{2}}=0.\label{eq:SB_SP_eq6}
\end{eqnarray}
\end{widetext}

\noindent where the symbol $f_{T}$ stands for the Fermi function of a single-particle matrix at temperature $T$ and we utilized the following matrix parameterizations:

\begin{align}
\Delta^p&=\sum_s d^p_s\,^t h_s,\label{eq:Deltap_param}\\ 
\lambda^c&=\sum_s l^c_s h_s,\label{eq:lamc_param}\\
\lambda&=\sum_s l_s h_s,\label{eq:lam_param}\\
R&=\sum_s r_s h_s\label{eq:R_param},
\end{align}

\noindent where the set of matrices $h_s$ are an orthonormal basis of the space of Hermitian matrices (with respect to the canonical trace inner product). The parameters $d^p_s$, $l^c_s$ and $l_s$ are real, while $r_s$ is complex.
The RISB saddle-point equations can be solved as follows:



\begin{enumerate}

\item Starting with an initial guess of $R$ and $\lambda$, compute
$\Delta^{p}$ from Eq. \eqref{eq:SB_SP_eq1}.

\item From $\Delta^{p}$, calculate $\mathcal{D}$ from Eq. \eqref{eq:SB_SP_eq2}.
\item With $\mathcal{D}$ and $\Delta^p$, compute
$\lambda^{c}$ from Eq. \eqref{eq:SB_SP_eq3}.

\item From
$\mathcal{D}$ and $\lambda^{c}$, construct $H_{\text{emb}}$ from Eq. \eqref{eq:Hemb} and 
calculate its ground state 
$|\Phi\rangle$.

\item From $|\Phi\rangle$ and $\Delta^p$, calculate 
Eqs.~\eqref{eq:SB_SP_eq5} and \eqref{eq:SB_SP_eq6} and utilize quasi-Newton methods to estimate the new $R$ and $\lambda$.

\item The convergence is achieved if Eqs.~\eqref{eq:SB_SP_eq5} and ~\eqref{eq:SB_SP_eq6} are satisfied.  Otherwise, continue the root searching with the new $R$ and $\lambda$.

\end{enumerate}

\begin{figure}
\begin{centering}
\includegraphics[scale=0.15]{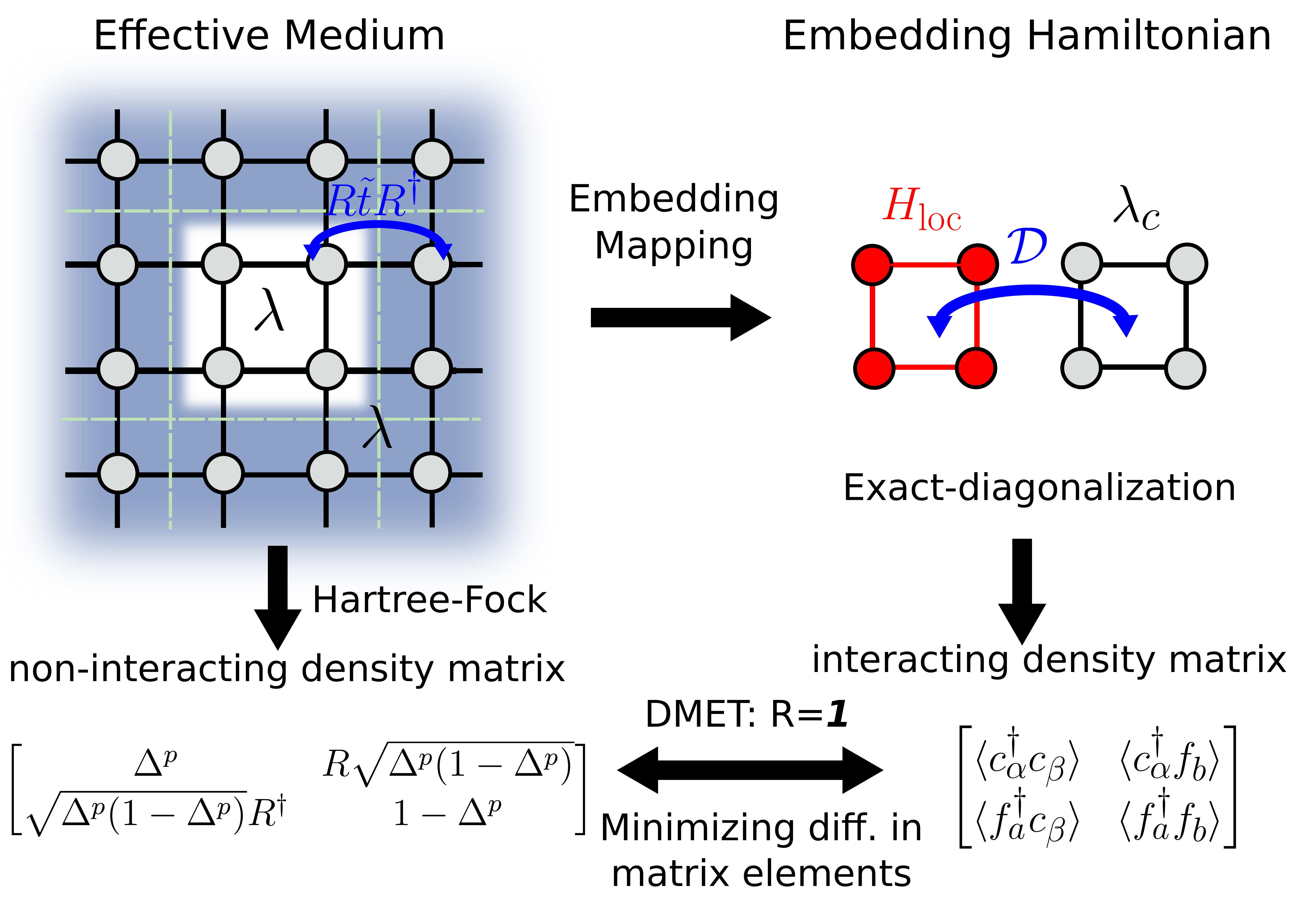}
\par\end{centering}
\caption{Schematic representation of the RISB and DMET algorithm. The black boxes denote the extra constraints for the DMET algorithm.
\label{fig:Schematic}}
\end{figure}

\noindent This structure is summarized schematically in Fig. \ref{fig:Schematic}.

Note that the Lagrange function [Eq. \ref{eq:emb_action}] evaluated for the converged parameters reduces to:

\begin{eqnarray}
&& E=\sum_{\mathbf{k}\in \text{RBZ}} \sum_{ab} \big[ R\tilde{\varepsilon}_{\mathbf{k}}R^\dagger f_T(R\tilde{\varepsilon}_{\mathbf{k}}R^\dagger+\lambda)\big]_{ab}
\nonumber\\
&& \qquad \qquad+\sum_i\langle\Phi|H_{i,loc}\big[c_{i\alpha}^\dagger,c_{i\alpha}\big]|\Phi\rangle,\label{eq:RISB_ene}
\end{eqnarray}

\noindent which is the total energy of the system.~\cite{Lanata2016}
It can be straightforwardly verified that,
as long as Eqs.~\eqref{eq:SB_SP_eq1}-\eqref{eq:SB_SP_eq6} are satisfied,
the total energy can be equivalently expressed also as follows:


\begin{equation}
E=\sum_i\langle\Phi|\sum_{\alpha a}(D_{\alpha a}c^\dagger_\alpha f_a) + H_{i,loc}[\{c^\dagger_\alpha c_\alpha\}]|\Phi\rangle\,.
\label{eq:DMET_ene}
\end{equation}



\subsection{Density matrix embedding theory}\label{dmet-algorithm}

The self-consistency conditions determining the parameters of $H_{\text{emb}}$ and $H_{\text{eff}}$ in DMET can be formulated as follows:\cite{Thomas_DMET_RISB}

\begin{widetext}
\begin{eqnarray}
&&\Delta_{ab}^{p}=\frac{N_c}{\mathcal{N}}\sum_{\mathbf{k}\in \text{RBZ}}\big[f_T(\tilde{\varepsilon}_{\mathbf{k}}+\lambda)\big]_{ba},\label{eq:DMET_eq1}\\
&&\big[\Delta^{p}(1-\Delta^{p})\big]^{1/2}_{ac}\mathcal{D}_{c\alpha}=\frac{N_c}{\mathcal{N}}\sum_{\mathbf{k}\in \text{RBZ}}\big[\tilde{\varepsilon}_{\mathbf{k}}f_T(\tilde{\varepsilon}_{\mathbf{k}}+\lambda)\big]_{\alpha a},\label{eq:DMET_eq2}\\
&&\sum_{cb}\frac{\partial}{\partial d^p_s}\big[\Delta^p(1-\Delta^p)\big]^{\frac{1}{2}}_{cb}\big[\mathcal{D}\big]_{bc}+\text{c.c}+\big[l+l^c\big]_s=0,\label{eq:DMET_eq3}\\
&&H_{\text{emb}}|\Phi\rangle=E^{c}|\Phi\rangle,\label{eq:DMET_eq4}\\
&&\Big[\mathcal{F}^{(1)}\Big]_{ab}\equiv\langle\Phi|f_{b}f_{a}^{\dagger}|\Phi\rangle-\Delta_{ab}^{p},\label{eq:DMET_root1}\\
&&\Big[\mathcal{F}^{(2)}\Big]_{\alpha a}\equiv\langle\Phi|c_{\alpha}^{\dagger}f_{a}|\Phi\rangle-\big[\Delta^{p}(1-\Delta^{p})\big]_{\alpha a}^{\frac{1}{2}},\label{eq:DMET_root2}\\
&&\Big[\mathcal{F}^{(3)}\Big]_{\alpha\beta}\equiv\langle\Phi|c_{\alpha}^{\dagger}c_{\beta}|\Phi\rangle-\Delta_{\alpha\beta}^{p},\label{eq:DMET_root3}\\
&&\lambda_{\text{min}}:=\underset{\lambda}{\text{argmin}}\ \big(\|\mathcal{F}^{(1)}\|_{\text{F}}+\|\mathcal{F}^{(2)}\|_{\text{F}}+\|\mathcal{F}^{(3)}\|_{\text{F}}\big)\,,\label{eq:DMET_norm}
\end{eqnarray}
\end{widetext}

\noindent where the symbol $\|...\|_{\text{F}}$ in Eq. \ref{eq:DMET_norm} indicates the Frobenius norm. Note that Eqs. \eqref{eq:DMET_eq1}-\eqref{eq:DMET_root2} are equivalent to Eqs. \eqref{eq:SB_SP_eq1}-\eqref{eq:SB_SP_eq6} with $R=\mathbf{1}$ and the constraint Eq. \eqref{eq:DMET_root3} was originally considered also in the Gutzwiller approximation (equivalent to RISB), but later was found to be unnecessary\cite{Fabrizio2007}. 



The DMET equations can be solved as follows, see Fig.~\ref{fig:Schematic}:
%


\begin{enumerate}

\item Starting with an initial guess of $\lambda$, calculate $\Delta^p$ using Eq. \eqref{eq:DMET_eq1}.

\item Compute $\mathcal{D}$ and $\lambda_c$ from Eq. \eqref{eq:DMET_eq2} and Eq. \eqref{eq:DMET_eq3} and construct the $H_{\text{emb}}$.

\item Compute the ground state $|\Phi\rangle$ and 
the corresponding single-particle density matrix, \emph{i.e.}: $\langle\Phi| f_b f_a^\dagger |\Phi\rangle$, $\langle\Phi| c_{\alpha}^\dagger f_a |\Phi\rangle$ and $\langle\Phi|c_{\alpha}^\dagger c_{\beta} |\Phi\rangle $.

\item 
From $\langle\Phi| f_b f_a^\dagger |\Phi\rangle$, $\langle\Phi| c_{\alpha}^\dagger f_a |\Phi\rangle$ and $\langle\Phi|c_{\alpha}^\dagger c_{\beta} |\Phi\rangle$, determine the entries of $\lambda_{\text{min}}$ that minimize Eq. \ref{eq:DMET_norm}\cite{Zheng_thesis} (note that such a minimum is generally larger than zero in interacting systems \cite{Knizia2012,Thomas_DMET_RISB}). 

\item Iterate until $\lambda_{\text{min}}$ is converged.

\end{enumerate}

\noindent A quasi-Newton method \cite{DIIS} is usually utilized to accelerate the convergence of DMET iteration. 
Once convergence is reached, the DMET total energy is computed from Eq. \eqref{eq:DMET_ene}.\cite{Knizia2012} 

\section{Results}


Here, we benchmark RISB and DMET with cluster sizes $N_c=1,\ 2,\ 4,\ 6$ on the Hubbard model with the nearest neighbor hopping in 1D and 2D (on a square lattice). The DMET calculations below are all performed utilizing the implementation outlined in Sec. \ref{dmet-algorithm}, featuring a modified RISB algorithm with mass renormalization weights set to unity. 
Our results are compared to the DMET data obtained in Refs. \onlinecite{Zheng2017} and \onlinecite{PRX_cluster_benchmark}. 

\begin{figure}[t]
\begin{centering}
\includegraphics[scale=0.33]{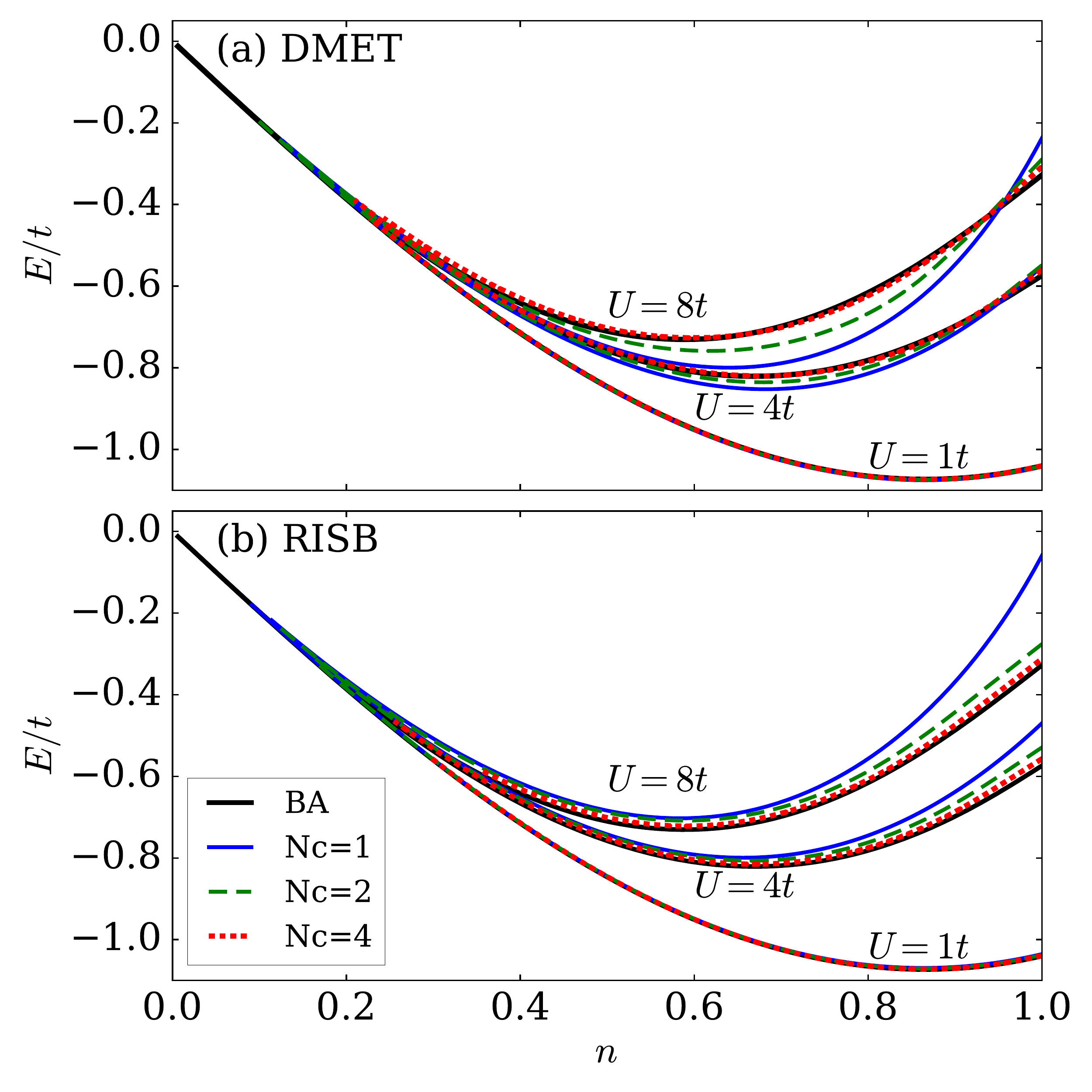}
\par\end{centering}
\caption{Energy $E/t$ for (a) DMET and (b) RISB as a function of occupancy $n$
in the 1D Hubbard model with the nearest neighbor hopping at $U=1t,\ 4t,\ 8t$ for cluster size
$N_{c}=1,\ 2,\ 4$, indicated by the blue solid, green dashed, and red dotted
lines, respectively. The solid black lines denote the results from BA. \label{fig:1D_energy}}
\end{figure}

\subsection{1D Hubbard model}


In Fig. \ref{fig:1D_energy} the DMET and RISB behaviors of the energies as a function of the occupation $n$ for $U\ =\ 1t,\ 4t,\ 8t$ with $N_c\ =\ 1,\ 2,\ 4$ are shown in comparison with the exact
Bethe Ansatz (BA)\cite{Bethe} solutions. Overall, the DMET and RISB approximations to the total energies are very similar for all cluster sizes, and both techniques reproduce the BA results with less than 2\% error already for $N_c= 4$. The only difference observed is that the DMET energies are slightly more accurate at half-filling, while the RISB energies are more accurate away from half-filling.

\begin{figure}[t]
\begin{centering}
\includegraphics[scale=0.33]{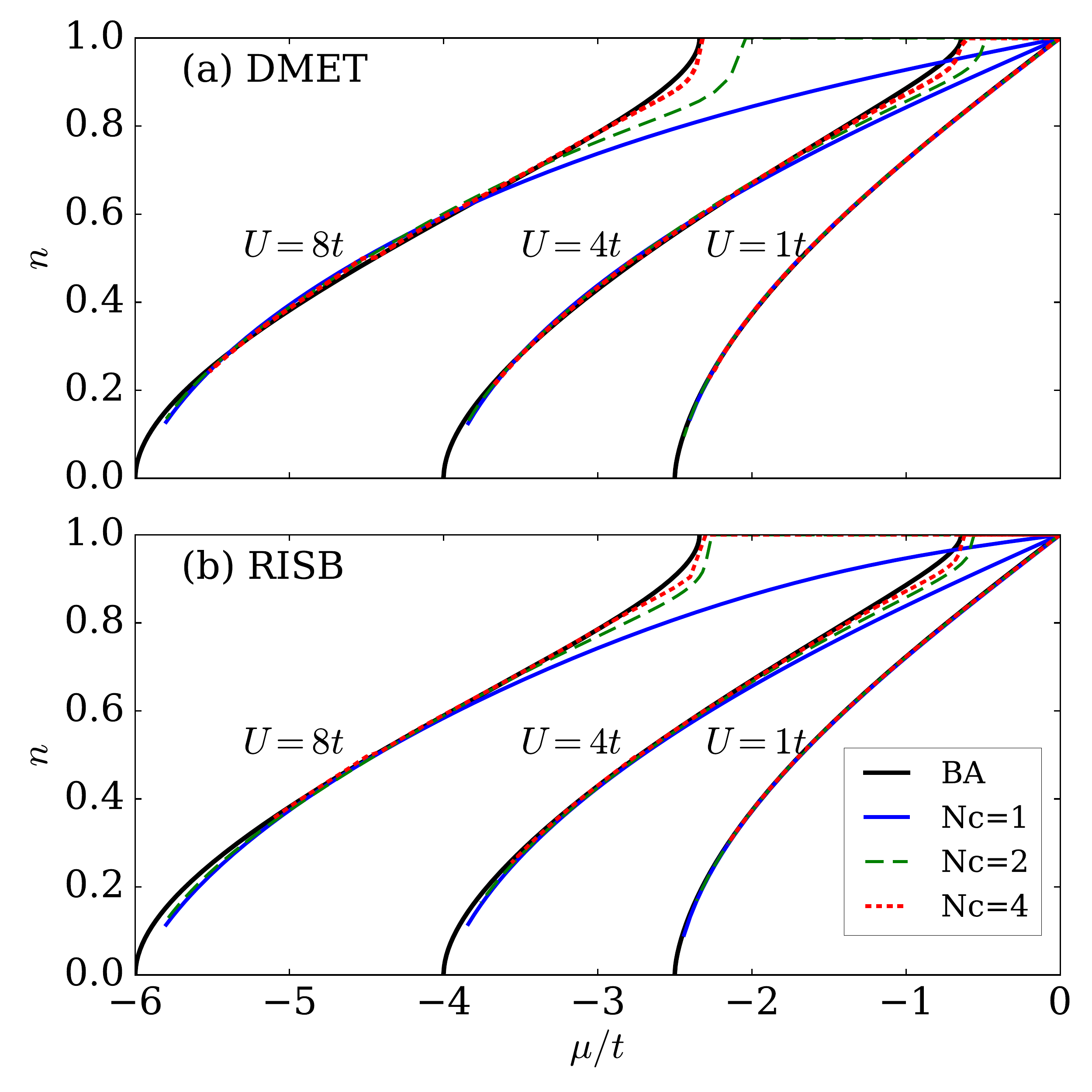}
\par\end{centering}
\caption{Occupancy $n$ for (a) DMET and (b) RISB as a function
of chemical potential $\mu$ in the 1D Hubbard model with the nearest neighbor hopping at $U=1t,\ 4t,\ 8t$
for cluster size $N_{c}=1,\ 2,\ 4$, indicated by the
blue solid, green dashed, and red dotted lines, respectively. The solid black lines denote
the results from BA.\label{fig:1D_gap}}
\end{figure}

In Figure \ref{fig:1D_gap} are shown the behaviors of the DMET and RISB occupancies $n$ as a function of the chemical potential $\mu$ for $U =\ 1t,\ 4t,\ 8t$ with $N_c =\ 1,\ 2,\ 4$, in comparison with the BA. The Mott insulating phase is characterized by a constant $n$ with compressibility $\frac{dn}{d\mu}=0$. At the Mott insulator-metal transition point $\mu_c$ the compressibility $\frac{dn}{d\mu}$ diverges\cite{capone2004}. In the metallic phase, $n$ decreases monotonically by decreasing $\mu$. We observe that both DMET and RISB capture the correct behavior for $N_c\geq2$. Moreover, RISB yields more accurate $n$ and $\mu_c$ at $N_c=2$. However, at $N_c=4$ both DMET and RISB predicts very precise occupancy and $\mu_c$ with less than 5\% error.

\vspace{2cm}


\begin{figure}[H]
\begin{centering}
\includegraphics[scale=0.33]{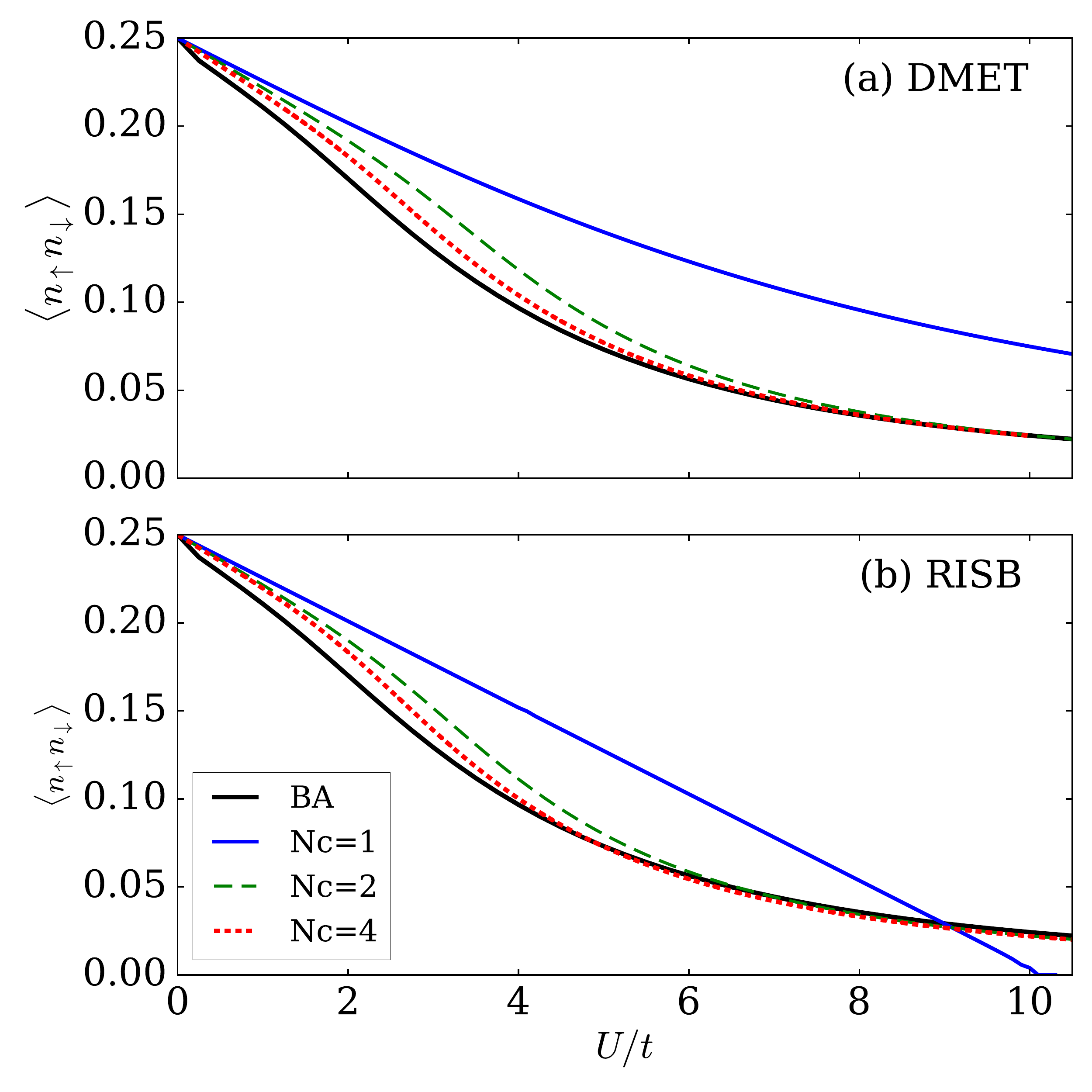}
\par\end{centering}
\caption{Double occupancy $\langle n_{\uparrow}n_{\downarrow}\rangle$
for (a) DMET and (b) RISB as a function of interaction $U$ in the half-filled 1D Hubbard with the nearest neighbor hopping for cluster size $N_{c}=1,\ 2,\ 4$,
indicated by the blue solid, green dashed, and red dotted lines, respectively. The solid
black lines denote the results from BA.\label{fig:1D_docc}}
\end{figure}

In Fig. \ref{fig:1D_docc} are shown the behaviors of the DMET and RISB double occupancies $\langle n_{\uparrow}n_{\downarrow}\rangle$ with $N_c\ =\ 1,\ 2,\ 4$, in comparison with the BA. At $N_{c}=1$ the DMET solutions are always metallic
for every $U$; consequently, the double occupancy deviates from the BA results at large $U$. On the other hand in RISB, the double occupancy vanishes at the critical point $U_{c}\sim10t$, \emph{i.e.}, the charge fluctuations are not captured in the Mott phase\cite{Brinkmann_Rice_1970}.
For $N_{c}=2$ both methods predict behaviors of $\langle n_{\uparrow}n_{\downarrow}\rangle$
that closely follow the BA values, although RISB is slightly more accurate. At $N_{c}=4$, both methods are very accurate with less than 7\% error compared
to BA. 

\begin{figure}[H]
\begin{centering}
\includegraphics[scale=0.36]{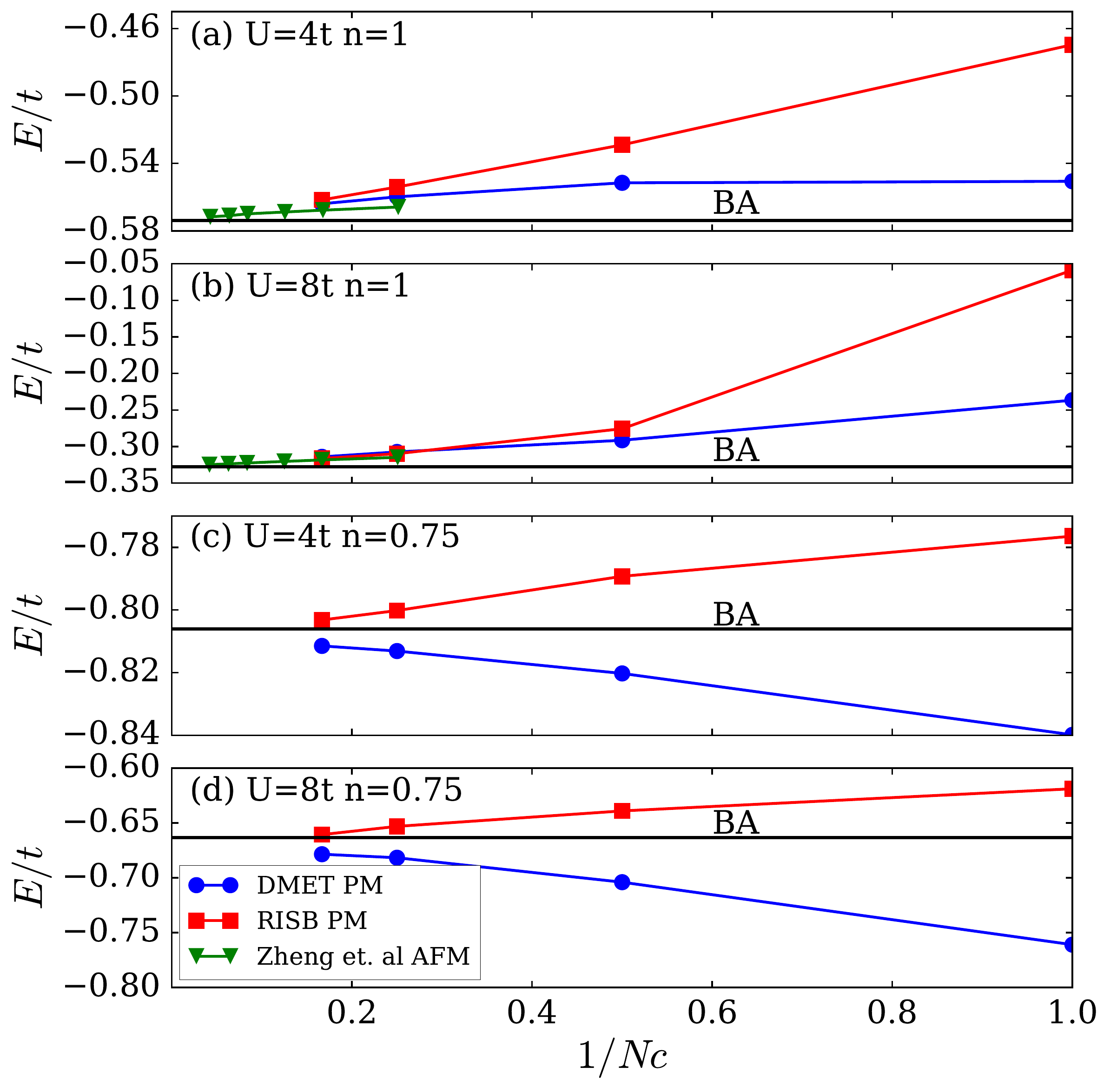}
\par\end{centering}
\caption{Energy $E/t$ as a function of inverse cluster size $1/N_{c}$ in the 1D Hubbard model with the nearest neighbor hopping
for (a) $U=4t$ and $n=1$, (b) $U=8t$ and $n=1$, (c) $U=4t$ and $n=0.75$, and (d) $U=8t$ and $n=0.75$. The blue circles correspond to the DMET values in our simulation. The red squares are our RISB results.
The green triangles are the data from Zheng et al. with antiferromagnetic
order \cite{Zheng2017}. The black solid lines are the results from BA.\label{fig:1D_ene_size}}
\end{figure}

We also analyze the convergence of the energy as a function of cluster
size at filling $n=1$ and $n=0.75$ with $U=4t$
and $U=8t$ for DMET and RISB as shown in Fig. \ref{fig:1D_ene_size}. DMET gives a better estimation for the ground-state energy at half-filling, while RISB yields more accurate energies at $n=0.75$. However, as the cluster size grows,
both methods converge to the BA value rapidly. 
Our results are consistent with the data extracted from Ref. \onlinecite{Zheng2017}, where an antiferromagnetic ground state was assumed (in 1D the ground state is non-magnetic).

\subsection{2D Hubbard model}

\begin{figure}[H]
\begin{centering}
\includegraphics[scale=0.38]{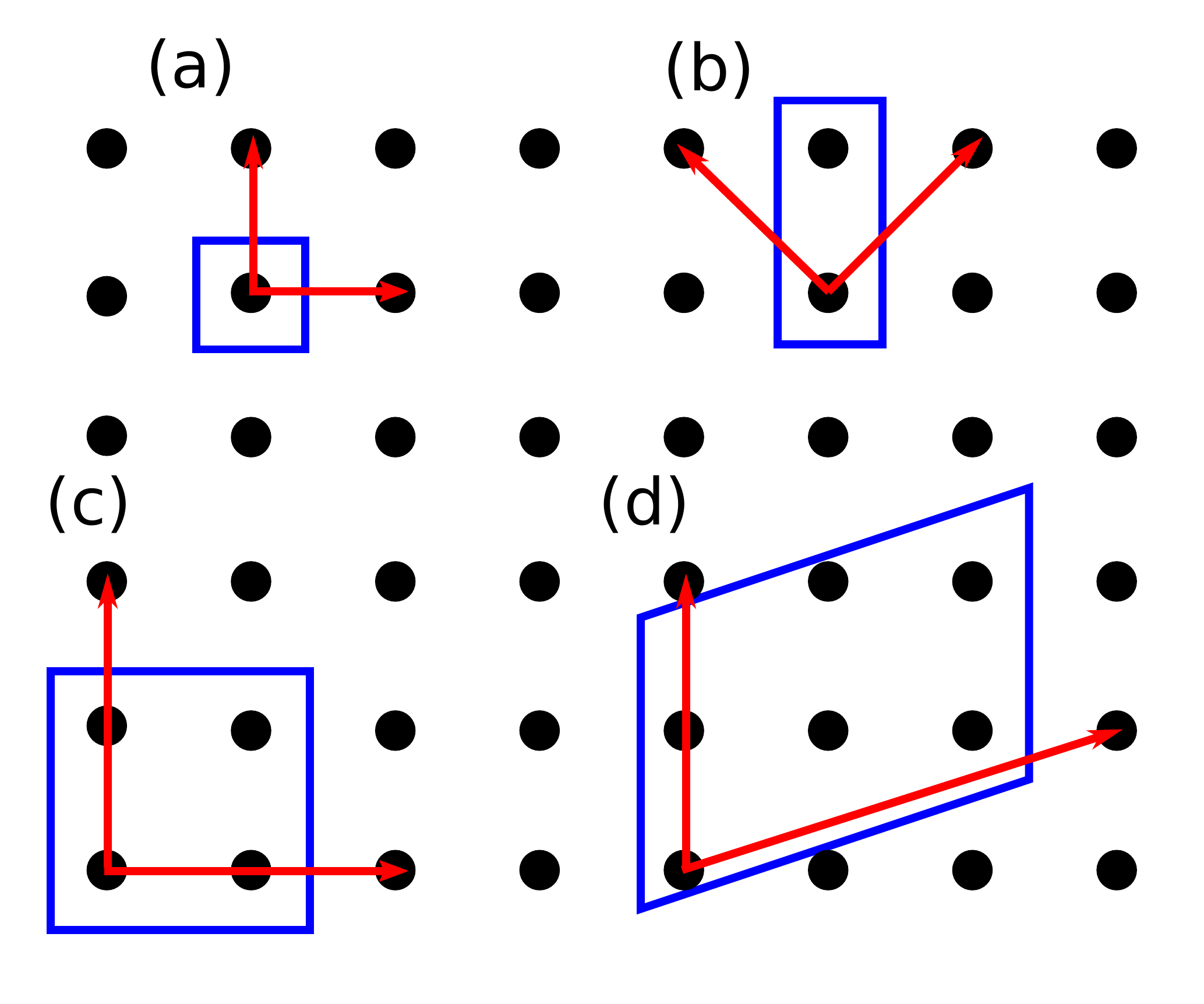}
\par\end{centering}
\caption{Clusters with sizes (a) $N_{c}=1$, (b) $N_{c}=2$, (c) $N_{c}=4$,
and (d) $N_{c}=6$, used in our simulation. The red arrows indicate the lattice vectors. The blue lines delimit the unit cells. \label{fig:2D_clusteres}}
\end{figure}

\begin{figure}[t]
\begin{centering}
\includegraphics[scale=0.35]{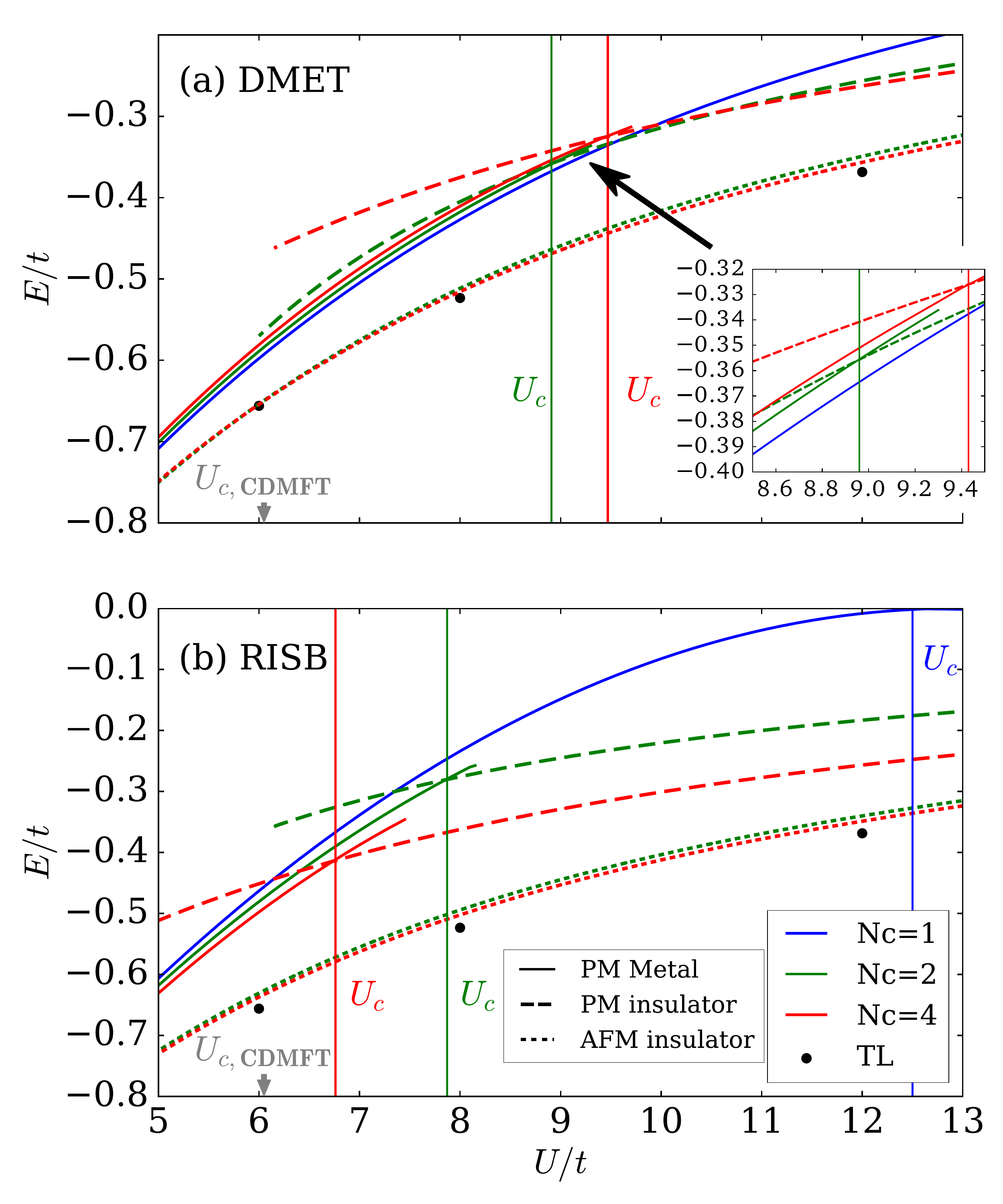}
\par\end{centering}
\caption{Energy $E/t$ for (a) DMET and (b) RISB as a function of interaction
$U$ in the half-filled 2D Hubbard model on a square lattice with the nearest neighbor hopping at cluster size, $N_{c}=1,\ 2,\ 4$,
indicated by the blue, green, and red line, respectively. The solid,
dashed, and dotted lines represent the PM metal, PM insulator,
and AFM solutions, respectively. The critical interaction $U_c$ is indicated by the verticle line. The black solid circles indicate the results in the TL from Ref. \onlinecite{Zheng2017} and \onlinecite{PRX_cluster_benchmark}. The grey arrow indicates the $U_c$ from Cellular-DMFT with $N_c=4$ in Ref. \onlinecite{Park_CDMFT}. The inset of (a) shows the magnified plot around $U_c$. \label{fig:2D_ene}}
\end{figure}

Here we investigate the behaviors of the RISB and DMET
solutions of the 2D Hubbard model on a square lattice with cluster sizes $N_{c}=1,\ 2,\ 4,\ 6$, see Fig. \ref{fig:2D_clusteres}. These geometries are chosen so that the antiferromagnetic (AFM) ground state can be reproduced for $N_c\geq2$ and that the paramagnetic (PM) and the AFM energetics can be compared on the same footing.


\begin{table*}
\begin{centering}
\begin{tabular}{|c|>{\centering}p{1.25cm}|>{\centering}p{1.25cm}|>{\centering}p{1.2cm}|>{\centering}p{1.2cm}|>{\centering}p{1.2cm}|>{\centering}p{1.2cm}|>{\centering}p{2.4cm}|>{\centering}p{2.4cm}|}
\hline 
 & \multicolumn{2}{c|}{$N_{c}=2$} & \multicolumn{2}{c|}{$N_{c}=4$} & \multicolumn{2}{c|}{$N_{c}=6$} &  $N_{c}=4$ Ref. \onlinecite{Zheng2017} & TL Ref. \onlinecite{PRX_cluster_benchmark}\tabularnewline
\hline 
Method & DMET & RISB & DMET & RISB & DMET & RISB & DMET & DMET\tabularnewline
\hline 
$U/t=2$ & -1.1804 & -1.1673 & -1.1790 & -1.1693 & -1.1790 & -1.1704 & -1.179 & -1.1764\tabularnewline
\hline 
$U/t=4$ & -0.8681 & -0.8428 & -0.8654 & -0.8459 & -0.8658 & -0.8472 & -0.863 & -0.8604\tabularnewline
\hline 
$U/t=6$ & -0.6541 & -0.6306 & -0.6545 & -0.6362 & -0.6553 & -0.6376 & -0.652 & -0.6562\tabularnewline
\hline 
$U/t=8$ & -0.5115 & -0.4942 & -0.5155 & -0.5023 & -0.5157 & -0.5100 & - & -0.5234\tabularnewline
\hline 
$U/t=12$ & -0.3497 & -0.3400  & -0.3566 & -0.3487& -0.3563 & -0.3565 & - & -0.3685\tabularnewline
\hline 
\end{tabular}
\par\end{centering}
\caption{Energy $E/t$ for DMET and RISB in the AFM phase of the 2D Hubbard model at half-filled $n=1$ with the nearest neighbor hopping for $N_c=2,\ 4,\ 6$ at $U=2t,\ 4t,\ 6t,\ 8t,\ 12t$. The values in the last two columns are the soltions at $N_c=4$ and in the TL extracted from Ref. \onlinecite{Zheng2017} and \onlinecite{PRX_cluster_benchmark}. \label{tab:2D_afm_ene}}
\end{table*}

\begin{table*}
\begin{centering}
\begin{tabular}{|c|>{\centering}p{1.25cm}|>{\centering}p{1.25cm}|>{\centering}p{1.2cm}|>{\centering}p{1.2cm}|>{\centering}p{1.2cm}|>{\centering}p{1.2cm}|>{\centering}p{2.4cm}|}
\hline 
 & \multicolumn{2}{c|}{$N_{c}=2$} & \multicolumn{2}{c|}{$N_{c}=4$} & \multicolumn{2}{c|}{$N_{c}=6$} & TL Ref. \onlinecite{PRX_cluster_benchmark}\tabularnewline
\hline 
Method & DMET & RISB & DMET & RISB & DMET & RISB & DMET\tabularnewline
\hline 
$U/t=2$ & -1.312 & -1.300 & -1.309 & -1.302 & -1.310 & -1.302 & -1.306 \tabularnewline
\hline 
$U/t=4$ & -1.129 & -1.083 & -1.122 & -1.086 & -1.120 & -1.091 & -1.108 \tabularnewline
\hline 
$U/t=6$ & -1.015 & -0.927 & -1.002 & -0.938 & -1.002 & -0.942 & -0.977\tabularnewline
\hline 
$U/t=8$ & -0.950 & -0.823 & -0.932 & -0.838 & -0.923 & -0.846 & -0.880 \tabularnewline
\hline  
\end{tabular}
\par\end{centering}
\caption{Energy $E/t$ for DMET and RISB in the PM phase of the 2D Hubbard model at $n=0.8$ with the nearest neighbor hopping for $N_c=2,\ 4,\ 6$ at $U=2t,\ 4t,\ 6t,\ 8t$. The values in the last two columns are the solutions at $N_c=4$ and in the TL extracted from Ref. \onlinecite{PRX_cluster_benchmark}. \label{tab:2D_ene_dope}}
\end{table*}

In Fig. \ref{fig:2D_ene} are shown the behaviors of the DMET and RISB total energy $E$ as a function of the Hubbard interaction $U$ at half-filling $n=1$ in the PM metal, PM insulating and AFM insulating phase, with cluster sizes $N_c = 1,\ 2,\ 4$.

At $N_{c}=1$, DMET does not capture the Mott metal-insulator transition (MIT), \emph{i.e.}, it predicts a metallic solution for every value of U. On the other hand, RISB predicts a MIT at $U_c=12.6t$, where the total energy vanishes \cite{Brinkmann_Rice_1970}. For $N_{c}\geq2$, both methods capture a MIT, as indicated by the crossing of the PM metal and PM insulator energies. Moreover, the energies of the AFM solutions are lower than the PM solutions, consistently with previous studies \cite{Knizia2012}.

It is also interesting to see how $U_{c}$ varies with the cluster size. We observe that in DMET $U_{c}$ is almost independent of the cluster size, \emph{e.g.}, $U_c=8.95t$ for $N_c=2$ and $U_c=9.65t$ for $N_c=4$. On the other hand, in RISB $U_{c}$ decreases from $12.6t$ for $N_c=1$ to $6.4t$ for $N_c=4$ (which is very close to the CDMFT value $U_c=6.05t$ for the same cluster size \cite{Park_CDMFT}).

\begin{figure}[t]
\begin{centering}
\includegraphics[scale=0.34]{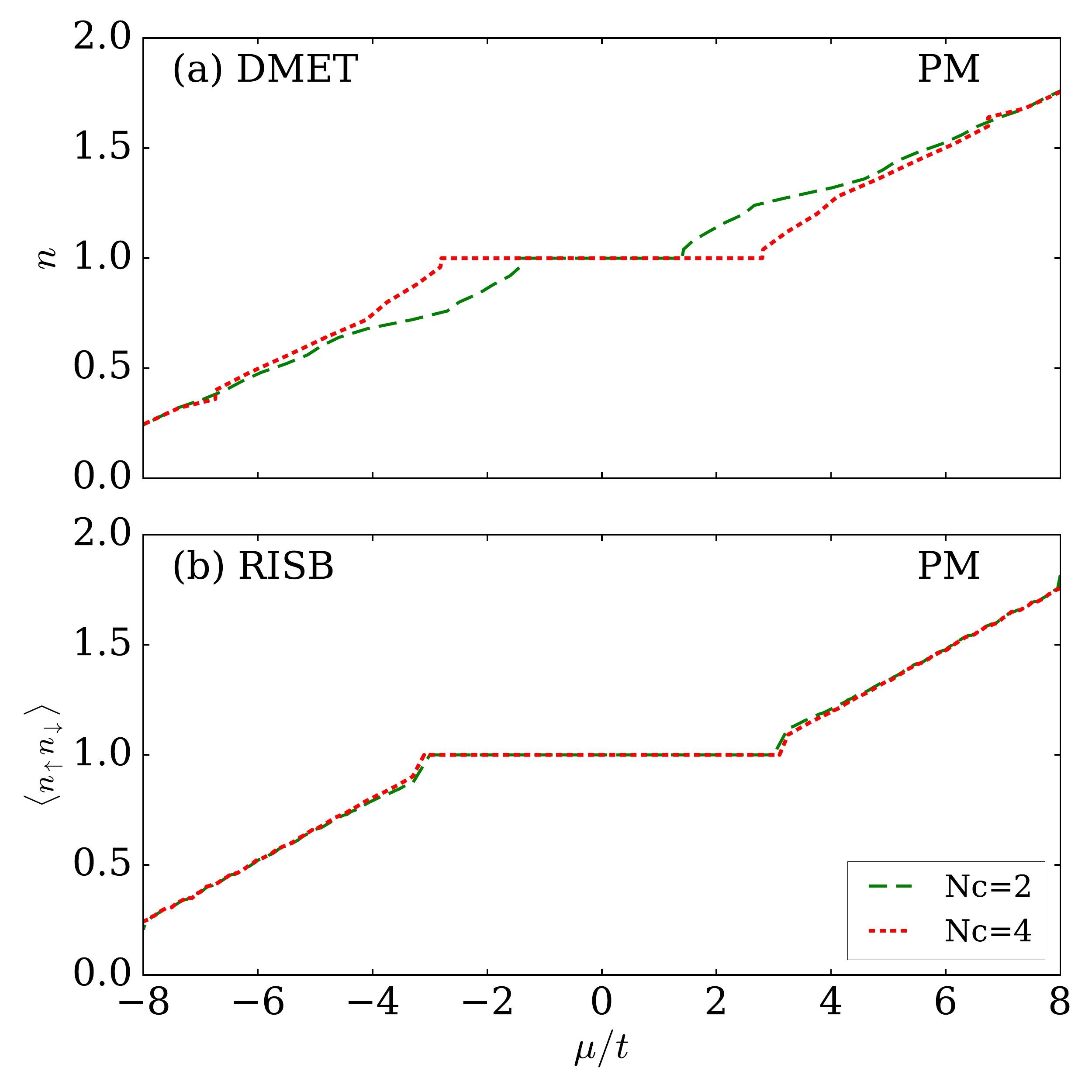}
\par\end{centering}
\caption{Occupancy $n$ as a function of chemical potential $\mu$ in the PM phase of the 2D Hubbard model on a square lattice with the nearest neighbor hopping
at $U=12t$ for cluster sizes $N_{c}=2\ \text{and}\ 4$, indicated by the green dashed and
red dotted line, respectively. 
\label{fig:2d_gap}}
\end{figure}

Figure \ref{fig:2d_gap} shows the DMET and RISB occupancy $n$ as a function of chemical potential $\mu$ at $U=12t$ with $N_c\ =\ 2,\ 4$. We observe that in DMET the difference in the occupancy and the $\mu_c$ between $N_c=2$ and $N_c=4$ is large, while in RISB, the discrepancy between the two cluster sizes is small (less than 3\% error). We conclude that RISB provides a slightly better description of the PM solutions.


\begin{table*}
\begin{centering}
\begin{tabular}{|c|>{\centering}p{1.25cm}|>{\centering}p{1.25cm}|>{\centering}p{1.2cm}|>{\centering}p{1.2cm}|>{\centering}p{1.2cm}|>{\centering}p{1.2cm}|>{\centering}p{2.4cm}|}
\hline 
 & \multicolumn{2}{c|}{$N_{c}=2$} & \multicolumn{2}{c|}{$N_{c}=4$} & \multicolumn{2}{c|}{$N_{c}=6$} & TL Ref. \onlinecite{PRX_cluster_benchmark}\tabularnewline
\hline 
Method & DMET & RISB & DMET & RISB & DMET & RISB & DMET\tabularnewline
\hline 
$U/t=2$ & 0.1937 & 0.1942 & 0.1934 & 0.1953 & 0.1935 & 0.1950 & 0.1913\tabularnewline
\hline 
$U/t=4$ & 0.1281 & 0.1314 & 0.1274 & 0.1300 & 0.1277 &  0.1300 & 0.1261\tabularnewline
\hline 
$U/t=6$ & 0.0819 & 0.0841 & 0.0815 & 0.0829 & 0.0816 & 0.0830 & 0.0810\tabularnewline
\hline 
$U/t=8$ & 0.0538 & 0.0548 & 0.0538 & 0.0542 & 0.0539 & 0.0541 & 0.0540\tabularnewline
\hline 
$U/t=12$ & 0.0268 & 0.0269  & 0.0272 & 0.0270 & 0.0272 & 0.0270 & 0.0278\tabularnewline
\hline 
\end{tabular}
\par\end{centering}
\caption{Double occupancy $\langle n_{\uparrow}n_{\downarrow} \rangle$ for DMET and RISB in the AFM phase of the half-filled 2D Hubbard model with the nearest neighbor hopping for $N_c=2,\ 4,\ 6$ at $U=2t,\ 4t,\ 6t,\ 8t,\ 12t$. The values in the last column are the solutions in the TL extracted from Ref. \onlinecite{PRX_cluster_benchmark}. \label{tab:2D_afm_docc}}
\end{table*}


\begin{table*}
\begin{centering}
\begin{tabular}{|c|>{\centering}p{1.25cm}|>{\centering}p{1.25cm}|>{\centering}p{1.2cm}|>{\centering}p{1.2cm}|>{\centering}p{1.2cm}|>{\centering}p{1.2cm}|>{\centering}p{2.4cm}|>{\centering}p{2.4cm}|}
\hline 
 & \multicolumn{2}{c|}{$N_{c}=2$} & \multicolumn{2}{c|}{$N_{c}=4$} & \multicolumn{2}{c|}{$N_{c}=6$} &  $N_{c}=4$ Ref. \onlinecite{Zheng2017} & TL Ref. \onlinecite{Zheng2017}\tabularnewline
\hline 
Method & DMET & RISB & DMET & RISB & DMET & RISB & DMET & DMET\tabularnewline
\hline 
$U/t=2$ & 0.161 & 0.158 & 0.155 & 0.147 & 0.151 & 0.143 & 0.152 & 0.115\tabularnewline
\hline 
$U/t=4$ & 0.304 & 0.293 & 0.298 & 0.289 & 0.296 & 0.288 & 0.299 & 0.226\tabularnewline
\hline 
$U/t=6$ & 0.382 & 0.376 & 0.368 & 0.368 & 0.367 & 0.365 & 0.372 & 0.275\tabularnewline
\hline 
\end{tabular}
\par\end{centering}
\caption{Staggered magnetic moment $m$ for DMET and RISB in the AFM phase of the half-filled 2D Hubbard model with the nearest neighbor hopping for $N_c=2,\ 4,\ 6$ at $U=2t,\ 4t,\ 6t$. The values in the last two columns are the solutions at $N_c=4$ and in the TL extracted from Ref. \onlinecite{Zheng2017}. \label{tab:2D_afm_m}}
\end{table*}


The ground-state energy predicted from DMET and RISB are shown in Tabs. \ref{tab:2D_afm_ene} and \ref{tab:2D_ene_dope} for $n=1$ AFM phase and $n=0.8$ PM phase, respectively, with various $U$ and $N_c$.
Our numerical values are compared to the DMET results at $N_c=4$ and in the TL in Refs. \onlinecite{PRX_cluster_benchmark} and \onlinecite{Zheng2017}, which are also shown as black solid dots in Fig. \ref{fig:2D_ene} at $n=1$. 

We observe that at half-filling $n=1$ DMET gives overall more accurate predictions to the ground-state energies in the AFM phase compared to the TL energies \cite{PRX_cluster_benchmark} (see Tab. \ref{tab:2D_afm_ene} and Fig. \ref{fig:2D_ene}). However, the discrepancy between the two methods is already small at $N_c=4$ (less than 3\% error). Away from half-filling ($n=0.8$), the ground-state energies predicted by RISB and DMET are equally accurate compared to the energies in the TL \cite{PRX_cluster_benchmark}. Our DMET results are consistent with previous studies \cite{Zheng2017,PRX_cluster_benchmark}.

The double occupancies $\langle n_{\uparrow} n_{\downarrow} \rangle$ at $n=1$ in the AFM phase with different $N_c$ and $U$ are shown in 
Tab. \ref{tab:2D_afm_docc}. DMET yields slightly more precise double 
occupancy at $N_c=2$ for smaller $U$ compared to the TL results \cite{PRX_cluster_benchmark}. However, for $N_c=4$, both methods obtained very accurate double occupancy close to the TL (less than 3\% error). 

In Tab. \ref{tab:2D_afm_m} we present the prediction of the AFM magnetic moment $m$ for both methods with different cluster sizes $N_c$ and $U$. Overall, we found the DMET and RISB magnetic moment are very similar, with RISB slightly closer to the TL\cite{PRX_cluster_benchmark}. 

\section{Conclusions}

We have performed comparative benchmark calculations of RISB and DMET on the 1D and 2D (square lattice) Hubbard model with cluster sizes ranging from $N_c=1$ to $6$. We found that the overall performances of the two methods are very similar. Small differences are observed only for small cluster sizes, where RISB generally predicts slightly more accurate Mott MIT critical points, magnetic moments, occupancies and double occupancies. The DMET ground-state energy is usually more accurate around half-filling, while the RISB ground-state energy is more precise away from half-filling.

Furthermore, we proposed an alternative implementation of DMET featuring a modified RISB algorithm with a unity mass renormalization matrix. 
This formalism 
paves the ways for many generalizations. For example, the DFT+RISB derived in Ref.  \onlinecite{Lanata2015} can now be readily transposed to DFT+DMET.
The non-equilibrium extensions of both methods are also available \cite{Schiro2010,Schiro2011,Mazza2017,Kretchmer2018}. A systematic way of improving the accuracy of RISB without breaking translational symmetry has been recently proposed by introducing auxiliary ``\emph{ghost}" degrees of freedom \cite{ghost_GA}, and similar ideas have been applied also within the DMET framework \cite{Booth2018}. Other possible directions may be to generalize DMET to finite-temperature \cite{Sandri_finite_T,Lanata_GA_finite_T,Mazza2017} or extending RISB to systems with electron-phonon interactions or inter-site electron-electron interactions \cite{DMET_e_phonon,DMET_e-p_Reinhard,PRX_H_chain_benchmark}.

\section{Acknowledgements}
T.-H. L. thanks G. Booth and Q. Chen for useful discussions on the DMET algorithm. Y. Y. thanks for the supports from BNL CMS center. T.-H. L, T. A., and G. K. were  supported by the Department of Energy under Grant No. DE-FG02-99ER45761. N. L. was supported by the VILLUM FONDEN via the Centre of Excellence for Dirac
Materials (Grant No. 11744). This work used the Extreme Science and Engineering Discovery Environment (XSEDE) funded by NSF under Grants No. TG-DMR170121.

\newpage

\bibliography{ref}

\begin{thebibliography}{48}%
\makeatletter
\providecommand \@ifxundefined [1]{%
 \@ifx{#1\undefined}
}%
\providecommand \@ifnum [1]{%
 \ifnum #1\expandafter \@firstoftwo
 \else \expandafter \@secondoftwo
 \fi
}%
\providecommand \@ifx [1]{%
 \ifx #1\expandafter \@firstoftwo
 \else \expandafter \@secondoftwo
 \fi
}%
\providecommand \natexlab [1]{#1}%
\providecommand \enquote  [1]{``#1''}%
\providecommand \bibnamefont  [1]{#1}%
\providecommand \bibfnamefont [1]{#1}%
\providecommand \citenamefont [1]{#1}%
\providecommand \href@noop [0]{\@secondoftwo}%
\providecommand \href [0]{\begingroup \@sanitize@url \@href}%
\providecommand \@href[1]{\@@startlink{#1}\@@href}%
\providecommand \@@href[1]{\endgroup#1\@@endlink}%
\providecommand \@sanitize@url [0]{\catcode `\\12\catcode `\$12\catcode
  `\&12\catcode `\#12\catcode `\^12\catcode `\_12\catcode `\%12\relax}%
\providecommand \@@startlink[1]{}%
\providecommand \@@endlink[0]{}%
\providecommand \url  [0]{\begingroup\@sanitize@url \@url }%
\providecommand \@url [1]{\endgroup\@href {#1}{\urlprefix }}%
\providecommand \urlprefix  [0]{URL }%
\providecommand \Eprint [0]{\href }%
\providecommand \doibase [0]{http://dx.doi.org/}%
\providecommand \selectlanguage [0]{\@gobble}%
\providecommand \bibinfo  [0]{\@secondoftwo}%
\providecommand \bibfield  [0]{\@secondoftwo}%
\providecommand \translation [1]{[#1]}%
\providecommand \BibitemOpen [0]{}%
\providecommand \bibitemStop [0]{}%
\providecommand \bibitemNoStop [0]{.\EOS\space}%
\providecommand \EOS [0]{\spacefactor3000\relax}%
\providecommand \BibitemShut  [1]{\csname bibitem#1\endcsname}%
\let\auto@bib@innerbib\@empty
\bibitem [{\citenamefont {Georges}\ \emph {et~al.}(1996)\citenamefont
  {Georges}, \citenamefont {Kotliar}, \citenamefont {Krauth},\ and\
  \citenamefont {Rozenberg}}]{Georges1996}%
  \BibitemOpen
  \bibfield  {author} {\bibinfo {author} {\bibfnamefont {A.}~\bibnamefont
  {Georges}}, \bibinfo {author} {\bibfnamefont {G.}~\bibnamefont {Kotliar}},
  \bibinfo {author} {\bibfnamefont {W.}~\bibnamefont {Krauth}}, \ and\ \bibinfo
  {author} {\bibfnamefont {M.~J.}\ \bibnamefont {Rozenberg}},\ }\href {\doibase
  10.1103/RevModPhys.68.13} {\bibfield  {journal} {\bibinfo  {journal} {Reviews
  of Modern Physics}\ }\textbf {\bibinfo {volume} {68}},\ \bibinfo {pages} {13}
  (\bibinfo {year} {1996})}\BibitemShut {NoStop}%
\bibitem [{\citenamefont {Knizia}\ and\ \citenamefont
  {Chan}(2012)}]{Knizia2012}%
  \BibitemOpen
  \bibfield  {author} {\bibinfo {author} {\bibfnamefont {G.}~\bibnamefont
  {Knizia}}\ and\ \bibinfo {author} {\bibfnamefont {G.~K.-L.}\ \bibnamefont
  {Chan}},\ }\href {\doibase 10.1103/PhysRevLett.109.186404} {\bibfield
  {journal} {\bibinfo  {journal} {Physical Review Letters}\ }\textbf {\bibinfo
  {volume} {109}},\ \bibinfo {pages} {186404} (\bibinfo {year}
  {2012})}\BibitemShut {NoStop}%
\bibitem [{\citenamefont {Kotliar}\ \emph {et~al.}(2006)\citenamefont
  {Kotliar}, \citenamefont {Savrasov}, \citenamefont {Haule}, \citenamefont
  {Oudovenko}, \citenamefont {Parcollet},\ and\ \citenamefont
  {Marianetti}}]{LDA_DMFT_RMP}%
  \BibitemOpen
  \bibfield  {author} {\bibinfo {author} {\bibfnamefont {G.}~\bibnamefont
  {Kotliar}}, \bibinfo {author} {\bibfnamefont {S.~Y.}\ \bibnamefont
  {Savrasov}}, \bibinfo {author} {\bibfnamefont {K.}~\bibnamefont {Haule}},
  \bibinfo {author} {\bibfnamefont {V.~S.}\ \bibnamefont {Oudovenko}}, \bibinfo
  {author} {\bibfnamefont {O.}~\bibnamefont {Parcollet}}, \ and\ \bibinfo
  {author} {\bibfnamefont {C.~A.}\ \bibnamefont {Marianetti}},\ }\href
  {\doibase 10.1103/RevModPhys.78.865} {\bibfield  {journal} {\bibinfo
  {journal} {Rev. Mod. Phys.}\ }\textbf {\bibinfo {volume} {78}},\ \bibinfo
  {pages} {865} (\bibinfo {year} {2006})}\BibitemShut {NoStop}%
\bibitem [{\citenamefont {Maier}\ \emph {et~al.}(2005)\citenamefont {Maier},
  \citenamefont {Jarrell}, \citenamefont {Pruschke},\ and\ \citenamefont
  {Hettler}}]{Maier2005a}%
  \BibitemOpen
  \bibfield  {author} {\bibinfo {author} {\bibfnamefont {T.~A.}\ \bibnamefont
  {Maier}}, \bibinfo {author} {\bibfnamefont {M.}~\bibnamefont {Jarrell}},
  \bibinfo {author} {\bibfnamefont {T.}~\bibnamefont {Pruschke}}, \ and\
  \bibinfo {author} {\bibfnamefont {M.~H.}\ \bibnamefont {Hettler}},\ }\href
  {\doibase 10.1103/RevModPhys.77.1027} {\bibfield  {journal} {\bibinfo
  {journal} {Reviews of Modern Physics}\ }\textbf {\bibinfo {volume} {77}},\
  \bibinfo {pages} {1027} (\bibinfo {year} {2005})}\BibitemShut {NoStop}%
\bibitem [{\citenamefont {Hettler}\ \emph {et~al.}(1998)\citenamefont
  {Hettler}, \citenamefont {Tahvildar-Zadeh}, \citenamefont {Jarrell},
  \citenamefont {Pruschke},\ and\ \citenamefont {Krishnamurthy}}]{Hettler1998}%
  \BibitemOpen
  \bibfield  {author} {\bibinfo {author} {\bibfnamefont {M.~H.}\ \bibnamefont
  {Hettler}}, \bibinfo {author} {\bibfnamefont {A.~N.}\ \bibnamefont
  {Tahvildar-Zadeh}}, \bibinfo {author} {\bibfnamefont {M.}~\bibnamefont
  {Jarrell}}, \bibinfo {author} {\bibfnamefont {T.}~\bibnamefont {Pruschke}}, \
  and\ \bibinfo {author} {\bibfnamefont {H.~R.}\ \bibnamefont
  {Krishnamurthy}},\ }\href {\doibase 10.1103/PhysRevB.58.R7475} {\bibfield
  {journal} {\bibinfo  {journal} {Physical Review B}\ }\textbf {\bibinfo
  {volume} {58}},\ \bibinfo {pages} {R7475} (\bibinfo {year}
  {1998})}\BibitemShut {NoStop}%
\bibitem [{\citenamefont {Hettler}\ \emph {et~al.}(1999)\citenamefont
  {Hettler}, \citenamefont {Mukherjee}, \citenamefont {Jarrell},\ and\
  \citenamefont {Krishnamurthy}}]{Hettler1999}%
  \BibitemOpen
  \bibfield  {author} {\bibinfo {author} {\bibfnamefont {M.~H.}\ \bibnamefont
  {Hettler}}, \bibinfo {author} {\bibfnamefont {M.}~\bibnamefont {Mukherjee}},
  \bibinfo {author} {\bibfnamefont {M.}~\bibnamefont {Jarrell}}, \ and\
  \bibinfo {author} {\bibfnamefont {H.~R.}\ \bibnamefont {Krishnamurthy}},\
  }\href {\doibase 10.1103/PhysRevB.61.12739} {\bibfield  {journal} {\bibinfo
  {journal} {Physical Review B}\ }\textbf {\bibinfo {volume} {61}},\ \bibinfo
  {pages} {12739} (\bibinfo {year} {1999})}\BibitemShut {NoStop}%
\bibitem [{\citenamefont {Lichtenstein}\ and\ \citenamefont
  {Katsnelson}(2000)}]{Lichtenstein2000}%
  \BibitemOpen
  \bibfield  {author} {\bibinfo {author} {\bibfnamefont {A.~I.}\ \bibnamefont
  {Lichtenstein}}\ and\ \bibinfo {author} {\bibfnamefont {M.~I.}\ \bibnamefont
  {Katsnelson}},\ }\href {\doibase 10.1103/PhysRevB.62.R9283} {\bibfield
  {journal} {\bibinfo  {journal} {Physical Review B}\ }\textbf {\bibinfo
  {volume} {62}},\ \bibinfo {pages} {R9283} (\bibinfo {year}
  {2000})}\BibitemShut {NoStop}%
\bibitem [{\citenamefont {Kotliar}\ \emph {et~al.}(2001)\citenamefont
  {Kotliar}, \citenamefont {Savrasov}, \citenamefont {P{\'{a}}lsson},\ and\
  \citenamefont {Biroli}}]{Kotliar2001}%
  \BibitemOpen
  \bibfield  {author} {\bibinfo {author} {\bibfnamefont {G.}~\bibnamefont
  {Kotliar}}, \bibinfo {author} {\bibfnamefont {S.}~\bibnamefont {Savrasov}},
  \bibinfo {author} {\bibfnamefont {G.}~\bibnamefont {P{\'{a}}lsson}}, \ and\
  \bibinfo {author} {\bibfnamefont {G.}~\bibnamefont {Biroli}},\ }\href
  {\doibase 10.1103/PhysRevLett.87.186401} {\bibfield  {journal} {\bibinfo
  {journal} {Physical Review Letters}\ }\textbf {\bibinfo {volume} {87}},\
  \bibinfo {pages} {186401} (\bibinfo {year} {2001})}\BibitemShut {NoStop}%
\bibitem [{\citenamefont {Rohringer}\ \emph {et~al.}(2018)\citenamefont
  {Rohringer}, \citenamefont {Hafermann}, \citenamefont {Toschi}, \citenamefont
  {Katanin}, \citenamefont {Antipov}, \citenamefont {Katsnelson}, \citenamefont
  {Lichtenstein}, \citenamefont {Rubtsov},\ and\ \citenamefont
  {Held}}]{Rohringer2017}%
  \BibitemOpen
  \bibfield  {author} {\bibinfo {author} {\bibfnamefont {G.}~\bibnamefont
  {Rohringer}}, \bibinfo {author} {\bibfnamefont {H.}~\bibnamefont
  {Hafermann}}, \bibinfo {author} {\bibfnamefont {A.}~\bibnamefont {Toschi}},
  \bibinfo {author} {\bibfnamefont {A.~A.}\ \bibnamefont {Katanin}}, \bibinfo
  {author} {\bibfnamefont {A.~E.}\ \bibnamefont {Antipov}}, \bibinfo {author}
  {\bibfnamefont {M.~I.}\ \bibnamefont {Katsnelson}}, \bibinfo {author}
  {\bibfnamefont {A.~I.}\ \bibnamefont {Lichtenstein}}, \bibinfo {author}
  {\bibfnamefont {A.~N.}\ \bibnamefont {Rubtsov}}, \ and\ \bibinfo {author}
  {\bibfnamefont {K.}~\bibnamefont {Held}},\ }\href {\doibase
  10.1103/RevModPhys.90.025003} {\bibfield  {journal} {\bibinfo  {journal}
  {Rev. Mod. Phys.}\ }\textbf {\bibinfo {volume} {90}},\ \bibinfo {pages}
  {025003} (\bibinfo {year} {2018})}\BibitemShut {NoStop}%
\bibitem [{\citenamefont {Knizia}\ and\ \citenamefont
  {Chan}(2013)}]{Knizia2013}%
  \BibitemOpen
  \bibfield  {author} {\bibinfo {author} {\bibfnamefont {G.}~\bibnamefont
  {Knizia}}\ and\ \bibinfo {author} {\bibfnamefont {G.~K.-L.}\ \bibnamefont
  {Chan}},\ }\href {\doibase 10.1021/ct301044e} {\bibfield  {journal} {\bibinfo
   {journal} {Journal of Chemical Theory and Computation}\ }\textbf {\bibinfo
  {volume} {9}},\ \bibinfo {pages} {1428} (\bibinfo {year} {2013})}\BibitemShut
  {NoStop}%
\bibitem [{\citenamefont {Wouters}\ \emph {et~al.}(2016)\citenamefont
  {Wouters}, \citenamefont {Jim{\'{e}}nez-Hoyos}, \citenamefont {Sun},\ and\
  \citenamefont {Chan}}]{Wouters2016a}%
  \BibitemOpen
  \bibfield  {author} {\bibinfo {author} {\bibfnamefont {S.}~\bibnamefont
  {Wouters}}, \bibinfo {author} {\bibfnamefont {C.~A.}\ \bibnamefont
  {Jim{\'{e}}nez-Hoyos}}, \bibinfo {author} {\bibfnamefont {Q.}~\bibnamefont
  {Sun}}, \ and\ \bibinfo {author} {\bibfnamefont {G.~K.}\ \bibnamefont
  {Chan}},\ }\href {\doibase 10.1021/acs.jctc.6b00316} {\bibfield  {journal}
  {\bibinfo  {journal} {Journal of Chemical Theory and Computation}\ }\textbf
  {\bibinfo {volume} {12}},\ \bibinfo {pages} {2706} (\bibinfo {year}
  {2016})}\BibitemShut {NoStop}%
\bibitem [{\citenamefont {Zheng}\ and\ \citenamefont {Chan}(2016)}]{Zheng2016}%
  \BibitemOpen
  \bibfield  {author} {\bibinfo {author} {\bibfnamefont {B.-X.}\ \bibnamefont
  {Zheng}}\ and\ \bibinfo {author} {\bibfnamefont {G.~K.-L.}\ \bibnamefont
  {Chan}},\ }\href {\doibase 10.1103/PhysRevB.93.035126} {\bibfield  {journal}
  {\bibinfo  {journal} {Physical Review B}\ }\textbf {\bibinfo {volume} {93}},\
  \bibinfo {pages} {035126} (\bibinfo {year} {2016})}\BibitemShut {NoStop}%
\bibitem [{\citenamefont {Zheng}\ \emph
  {et~al.}(2017{\natexlab{a}})\citenamefont {Zheng}, \citenamefont {Kretchmer},
  \citenamefont {Shi}, \citenamefont {Zhang},\ and\ \citenamefont
  {Chan}}]{Zheng2017}%
  \BibitemOpen
  \bibfield  {author} {\bibinfo {author} {\bibfnamefont {B.-X.}\ \bibnamefont
  {Zheng}}, \bibinfo {author} {\bibfnamefont {J.~S.}\ \bibnamefont
  {Kretchmer}}, \bibinfo {author} {\bibfnamefont {H.}~\bibnamefont {Shi}},
  \bibinfo {author} {\bibfnamefont {S.}~\bibnamefont {Zhang}}, \ and\ \bibinfo
  {author} {\bibfnamefont {G.~K.-L.}\ \bibnamefont {Chan}},\ }\href {\doibase
  10.1103/PhysRevB.95.045103} {\bibfield  {journal} {\bibinfo  {journal}
  {Physical Review B}\ }\textbf {\bibinfo {volume} {95}},\ \bibinfo {pages}
  {045103} (\bibinfo {year} {2017}{\natexlab{a}})}\BibitemShut {NoStop}%
\bibitem [{\citenamefont {Zheng}\ \emph
  {et~al.}(2017{\natexlab{b}})\citenamefont {Zheng}, \citenamefont {Chung},
  \citenamefont {Corboz}, \citenamefont {Ehlers}, \citenamefont {Qin},
  \citenamefont {Noack}, \citenamefont {Shi}, \citenamefont {White},
  \citenamefont {Zhang},\ and\ \citenamefont {Chan}}]{Zheng_science_benchmark}%
  \BibitemOpen
  \bibfield  {author} {\bibinfo {author} {\bibfnamefont {B.-X.}\ \bibnamefont
  {Zheng}}, \bibinfo {author} {\bibfnamefont {C.-M.}\ \bibnamefont {Chung}},
  \bibinfo {author} {\bibfnamefont {P.}~\bibnamefont {Corboz}}, \bibinfo
  {author} {\bibfnamefont {G.}~\bibnamefont {Ehlers}}, \bibinfo {author}
  {\bibfnamefont {M.-P.}\ \bibnamefont {Qin}}, \bibinfo {author} {\bibfnamefont
  {R.~M.}\ \bibnamefont {Noack}}, \bibinfo {author} {\bibfnamefont
  {H.}~\bibnamefont {Shi}}, \bibinfo {author} {\bibfnamefont {S.~R.}\
  \bibnamefont {White}}, \bibinfo {author} {\bibfnamefont {S.}~\bibnamefont
  {Zhang}}, \ and\ \bibinfo {author} {\bibfnamefont {G.~K.-L.}\ \bibnamefont
  {Chan}},\ }\href {\doibase 10.1126/science.aam7127} {\bibfield  {journal}
  {\bibinfo  {journal} {Science}\ }\textbf {\bibinfo {volume} {358}},\ \bibinfo
  {pages} {1155} (\bibinfo {year} {2017}{\natexlab{b}})},\ \Eprint
  {http://arxiv.org/abs/http://science.sciencemag.org/content/358/6367/1155.full.pdf}
  {http://science.sciencemag.org/content/358/6367/1155.full.pdf} \BibitemShut
  {NoStop}%
\bibitem [{\citenamefont {LeBlanc}\ \emph {et~al.}(2015)\citenamefont
  {LeBlanc}, \citenamefont {Antipov}, \citenamefont {Becca}, \citenamefont
  {Bulik}, \citenamefont {Chan}, \citenamefont {Chung}, \citenamefont {Deng},
  \citenamefont {Ferrero}, \citenamefont {Henderson}, \citenamefont
  {Jim\'enez-Hoyos}, \citenamefont {Kozik}, \citenamefont {Liu}, \citenamefont
  {Millis}, \citenamefont {Prokof'ev}, \citenamefont {Qin}, \citenamefont
  {Scuseria}, \citenamefont {Shi}, \citenamefont {Svistunov}, \citenamefont
  {Tocchio}, \citenamefont {Tupitsyn}, \citenamefont {White}, \citenamefont
  {Zhang}, \citenamefont {Zheng}, \citenamefont {Zhu},\ and\ \citenamefont
  {Gull}}]{PRX_cluster_benchmark}%
  \BibitemOpen
  \bibfield  {author} {\bibinfo {author} {\bibfnamefont {J.~P.~F.}\
  \bibnamefont {LeBlanc}}, \bibinfo {author} {\bibfnamefont {A.~E.}\
  \bibnamefont {Antipov}}, \bibinfo {author} {\bibfnamefont {F.}~\bibnamefont
  {Becca}}, \bibinfo {author} {\bibfnamefont {I.~W.}\ \bibnamefont {Bulik}},
  \bibinfo {author} {\bibfnamefont {G.~K.-L.}\ \bibnamefont {Chan}}, \bibinfo
  {author} {\bibfnamefont {C.-M.}\ \bibnamefont {Chung}}, \bibinfo {author}
  {\bibfnamefont {Y.}~\bibnamefont {Deng}}, \bibinfo {author} {\bibfnamefont
  {M.}~\bibnamefont {Ferrero}}, \bibinfo {author} {\bibfnamefont {T.~M.}\
  \bibnamefont {Henderson}}, \bibinfo {author} {\bibfnamefont {C.~A.}\
  \bibnamefont {Jim\'enez-Hoyos}}, \bibinfo {author} {\bibfnamefont
  {E.}~\bibnamefont {Kozik}}, \bibinfo {author} {\bibfnamefont {X.-W.}\
  \bibnamefont {Liu}}, \bibinfo {author} {\bibfnamefont {A.~J.}\ \bibnamefont
  {Millis}}, \bibinfo {author} {\bibfnamefont {N.~V.}\ \bibnamefont
  {Prokof'ev}}, \bibinfo {author} {\bibfnamefont {M.}~\bibnamefont {Qin}},
  \bibinfo {author} {\bibfnamefont {G.~E.}\ \bibnamefont {Scuseria}}, \bibinfo
  {author} {\bibfnamefont {H.}~\bibnamefont {Shi}}, \bibinfo {author}
  {\bibfnamefont {B.~V.}\ \bibnamefont {Svistunov}}, \bibinfo {author}
  {\bibfnamefont {L.~F.}\ \bibnamefont {Tocchio}}, \bibinfo {author}
  {\bibfnamefont {I.~S.}\ \bibnamefont {Tupitsyn}}, \bibinfo {author}
  {\bibfnamefont {S.~R.}\ \bibnamefont {White}}, \bibinfo {author}
  {\bibfnamefont {S.}~\bibnamefont {Zhang}}, \bibinfo {author} {\bibfnamefont
  {B.-X.}\ \bibnamefont {Zheng}}, \bibinfo {author} {\bibfnamefont
  {Z.}~\bibnamefont {Zhu}}, \ and\ \bibinfo {author} {\bibfnamefont
  {E.}~\bibnamefont {Gull}} (\bibinfo {collaboration} {Simons Collaboration on
  the Many-Electron Problem}),\ }\href {\doibase 10.1103/PhysRevX.5.041041}
  {\bibfield  {journal} {\bibinfo  {journal} {Phys. Rev. X}\ }\textbf {\bibinfo
  {volume} {5}},\ \bibinfo {pages} {041041} (\bibinfo {year}
  {2015})}\BibitemShut {NoStop}%
\bibitem [{\citenamefont {Motta}\ \emph {et~al.}(2017)\citenamefont {Motta},
  \citenamefont {Ceperley}, \citenamefont {Chan}, \citenamefont {Gomez},
  \citenamefont {Gull}, \citenamefont {Guo}, \citenamefont {Jim\'enez-Hoyos},
  \citenamefont {Lan}, \citenamefont {Li}, \citenamefont {Ma}, \citenamefont
  {Millis}, \citenamefont {Prokof'ev}, \citenamefont {Ray}, \citenamefont
  {Scuseria}, \citenamefont {Sorella}, \citenamefont {Stoudenmire},
  \citenamefont {Sun}, \citenamefont {Tupitsyn}, \citenamefont {White},
  \citenamefont {Zgid},\ and\ \citenamefont {Zhang}}]{PRX_H_chain_benchmark}%
  \BibitemOpen
  \bibfield  {author} {\bibinfo {author} {\bibfnamefont {M.}~\bibnamefont
  {Motta}}, \bibinfo {author} {\bibfnamefont {D.~M.}\ \bibnamefont {Ceperley}},
  \bibinfo {author} {\bibfnamefont {G.~K.-L.}\ \bibnamefont {Chan}}, \bibinfo
  {author} {\bibfnamefont {J.~A.}\ \bibnamefont {Gomez}}, \bibinfo {author}
  {\bibfnamefont {E.}~\bibnamefont {Gull}}, \bibinfo {author} {\bibfnamefont
  {S.}~\bibnamefont {Guo}}, \bibinfo {author} {\bibfnamefont {C.~A.}\
  \bibnamefont {Jim\'enez-Hoyos}}, \bibinfo {author} {\bibfnamefont {T.~N.}\
  \bibnamefont {Lan}}, \bibinfo {author} {\bibfnamefont {J.}~\bibnamefont
  {Li}}, \bibinfo {author} {\bibfnamefont {F.}~\bibnamefont {Ma}}, \bibinfo
  {author} {\bibfnamefont {A.~J.}\ \bibnamefont {Millis}}, \bibinfo {author}
  {\bibfnamefont {N.~V.}\ \bibnamefont {Prokof'ev}}, \bibinfo {author}
  {\bibfnamefont {U.}~\bibnamefont {Ray}}, \bibinfo {author} {\bibfnamefont
  {G.~E.}\ \bibnamefont {Scuseria}}, \bibinfo {author} {\bibfnamefont
  {S.}~\bibnamefont {Sorella}}, \bibinfo {author} {\bibfnamefont {E.~M.}\
  \bibnamefont {Stoudenmire}}, \bibinfo {author} {\bibfnamefont
  {Q.}~\bibnamefont {Sun}}, \bibinfo {author} {\bibfnamefont {I.~S.}\
  \bibnamefont {Tupitsyn}}, \bibinfo {author} {\bibfnamefont {S.~R.}\
  \bibnamefont {White}}, \bibinfo {author} {\bibfnamefont {D.}~\bibnamefont
  {Zgid}}, \ and\ \bibinfo {author} {\bibfnamefont {S.}~\bibnamefont {Zhang}}
  (\bibinfo {collaboration} {Simons Collaboration on the Many-Electron
  Problem}),\ }\href {\doibase 10.1103/PhysRevX.7.031059} {\bibfield  {journal}
  {\bibinfo  {journal} {Phys. Rev. X}\ }\textbf {\bibinfo {volume} {7}},\
  \bibinfo {pages} {031059} (\bibinfo {year} {2017})}\BibitemShut {NoStop}%
\bibitem [{\citenamefont {Fr{\'{e}}sard}\ and\ \citenamefont
  {W{\"{o}}lfle}(1992)}]{Fresard1992}%
  \BibitemOpen
  \bibfield  {author} {\bibinfo {author} {\bibfnamefont {R.}~\bibnamefont
  {Fr{\'{e}}sard}}\ and\ \bibinfo {author} {\bibfnamefont {P.}~\bibnamefont
  {W{\"{o}}lfle}},\ }\href {\doibase 10.1142/S0217979292000414} {\bibfield
  {journal} {\bibinfo  {journal} {International Journal of Modern Physics B}\
  }\textbf {\bibinfo {volume} {06}},\ \bibinfo {pages} {685} (\bibinfo {year}
  {1992})}\BibitemShut {NoStop}%
\bibitem [{\citenamefont {Lechermann}\ \emph {et~al.}(2007)\citenamefont
  {Lechermann}, \citenamefont {Georges}, \citenamefont {Kotliar},\ and\
  \citenamefont {Parcollet}}]{Lechermann2007}%
  \BibitemOpen
  \bibfield  {author} {\bibinfo {author} {\bibfnamefont {F.}~\bibnamefont
  {Lechermann}}, \bibinfo {author} {\bibfnamefont {A.}~\bibnamefont {Georges}},
  \bibinfo {author} {\bibfnamefont {G.}~\bibnamefont {Kotliar}}, \ and\
  \bibinfo {author} {\bibfnamefont {O.}~\bibnamefont {Parcollet}},\ }\href
  {\doibase 10.1103/PhysRevB.76.155102} {\bibfield  {journal} {\bibinfo
  {journal} {Physical Review B}\ }\textbf {\bibinfo {volume} {76}},\ \bibinfo
  {pages} {155102} (\bibinfo {year} {2007})}\BibitemShut {NoStop}%
\bibitem [{\citenamefont {Lanat{\`{a}}}\ \emph {et~al.}(2017)\citenamefont
  {Lanat{\`{a}}}, \citenamefont {Yao}, \citenamefont {Deng}, \citenamefont
  {Dobrosavljevi{\'{c}}},\ and\ \citenamefont {Kotliar}}]{Lanata2016}%
  \BibitemOpen
  \bibfield  {author} {\bibinfo {author} {\bibfnamefont {N.}~\bibnamefont
  {Lanat{\`{a}}}}, \bibinfo {author} {\bibfnamefont {Y.}~\bibnamefont {Yao}},
  \bibinfo {author} {\bibfnamefont {X.}~\bibnamefont {Deng}}, \bibinfo {author}
  {\bibfnamefont {V.}~\bibnamefont {Dobrosavljevi{\'{c}}}}, \ and\ \bibinfo
  {author} {\bibfnamefont {G.}~\bibnamefont {Kotliar}},\ }\href {\doibase
  10.1103/PhysRevLett.118.126401} {\bibfield  {journal} {\bibinfo  {journal}
  {Physical Review Letters}\ }\textbf {\bibinfo {volume} {118}},\ \bibinfo
  {pages} {126401} (\bibinfo {year} {2017})}\BibitemShut {NoStop}%
\bibitem [{\citenamefont {Kotliar}\ and\ \citenamefont
  {Ruckenstein}(1986)}]{Kotliar1986}%
  \BibitemOpen
  \bibfield  {author} {\bibinfo {author} {\bibfnamefont {G.}~\bibnamefont
  {Kotliar}}\ and\ \bibinfo {author} {\bibfnamefont {A.~E.}\ \bibnamefont
  {Ruckenstein}},\ }\href {\doibase 10.1103/PhysRevLett.57.1362} {\bibfield
  {journal} {\bibinfo  {journal} {Physical Review Letters}\ }\textbf {\bibinfo
  {volume} {57}},\ \bibinfo {pages} {1362} (\bibinfo {year}
  {1986})}\BibitemShut {NoStop}%
\bibitem [{\citenamefont {B{\"{u}}nemann}\ and\ \citenamefont
  {Gebhard}(2007)}]{Bunemann2007}%
  \BibitemOpen
  \bibfield  {author} {\bibinfo {author} {\bibfnamefont {J.}~\bibnamefont
  {B{\"{u}}nemann}}\ and\ \bibinfo {author} {\bibfnamefont {F.}~\bibnamefont
  {Gebhard}},\ }\href {\doibase 10.1103/PhysRevB.76.193104} {\bibfield
  {journal} {\bibinfo  {journal} {Physical Review B}\ }\textbf {\bibinfo
  {volume} {76}},\ \bibinfo {pages} {193104} (\bibinfo {year}
  {2007})}\BibitemShut {NoStop}%
\bibitem [{\citenamefont {Lanat\`a}\ \emph {et~al.}(2008)\citenamefont
  {Lanat\`a}, \citenamefont {Barone},\ and\ \citenamefont
  {Fabrizio}}]{Lanata_KLM}%
  \BibitemOpen
  \bibfield  {author} {\bibinfo {author} {\bibfnamefont {N.}~\bibnamefont
  {Lanat\`a}}, \bibinfo {author} {\bibfnamefont {P.}~\bibnamefont {Barone}}, \
  and\ \bibinfo {author} {\bibfnamefont {M.}~\bibnamefont {Fabrizio}},\ }\href
  {\doibase 10.1103/PhysRevB.78.155127} {\bibfield  {journal} {\bibinfo
  {journal} {Phys. Rev. B}\ }\textbf {\bibinfo {volume} {78}},\ \bibinfo
  {pages} {155127} (\bibinfo {year} {2008})}\BibitemShut {NoStop}%
\bibitem [{\citenamefont {Isidori}\ and\ \citenamefont
  {Capone}(2009)}]{Isidori2009}%
  \BibitemOpen
  \bibfield  {author} {\bibinfo {author} {\bibfnamefont {A.}~\bibnamefont
  {Isidori}}\ and\ \bibinfo {author} {\bibfnamefont {M.}~\bibnamefont
  {Capone}},\ }\href {\doibase 10.1103/PhysRevB.80.115120} {\bibfield
  {journal} {\bibinfo  {journal} {Physical Review B}\ }\textbf {\bibinfo
  {volume} {80}},\ \bibinfo {pages} {115120} (\bibinfo {year}
  {2009})}\BibitemShut {NoStop}%
\bibitem [{\citenamefont {Ferrero}\ \emph {et~al.}(2008)\citenamefont
  {Ferrero}, \citenamefont {Cornaglia}, \citenamefont {{De Leo}}, \citenamefont
  {Parcollet}, \citenamefont {Kotliar},\ and\ \citenamefont
  {Georges}}]{Ferrero2008}%
  \BibitemOpen
  \bibfield  {author} {\bibinfo {author} {\bibfnamefont {M.}~\bibnamefont
  {Ferrero}}, \bibinfo {author} {\bibfnamefont {P.~S.}\ \bibnamefont
  {Cornaglia}}, \bibinfo {author} {\bibfnamefont {L.}~\bibnamefont {{De Leo}}},
  \bibinfo {author} {\bibfnamefont {O.}~\bibnamefont {Parcollet}}, \bibinfo
  {author} {\bibfnamefont {G.}~\bibnamefont {Kotliar}}, \ and\ \bibinfo
  {author} {\bibfnamefont {A.}~\bibnamefont {Georges}},\ }\href {\doibase
  10.1209/0295-5075/85/57009} {\bibfield  {journal} {\bibinfo  {journal}
  {Europhysics Letters}\ }\textbf {\bibinfo {volume} {85}},\ \bibinfo {pages}
  {57009} (\bibinfo {year} {2008})}\BibitemShut {NoStop}%
\bibitem [{\citenamefont {Ferrero}\ \emph {et~al.}(2009)\citenamefont
  {Ferrero}, \citenamefont {Cornaglia}, \citenamefont {{De Leo}}, \citenamefont
  {Parcollet}, \citenamefont {Kotliar},\ and\ \citenamefont
  {Georges}}]{Ferrero2009}%
  \BibitemOpen
  \bibfield  {author} {\bibinfo {author} {\bibfnamefont {M.}~\bibnamefont
  {Ferrero}}, \bibinfo {author} {\bibfnamefont {P.}~\bibnamefont {Cornaglia}},
  \bibinfo {author} {\bibfnamefont {L.}~\bibnamefont {{De Leo}}}, \bibinfo
  {author} {\bibfnamefont {O.}~\bibnamefont {Parcollet}}, \bibinfo {author}
  {\bibfnamefont {G.}~\bibnamefont {Kotliar}}, \ and\ \bibinfo {author}
  {\bibfnamefont {A.}~\bibnamefont {Georges}},\ }\href {\doibase
  10.1103/PhysRevB.80.064501} {\bibfield  {journal} {\bibinfo  {journal}
  {Physical Review B}\ }\textbf {\bibinfo {volume} {80}},\ \bibinfo {pages}
  {064501} (\bibinfo {year} {2009})}\BibitemShut {NoStop}%
\bibitem [{\citenamefont {Mazin}\ \emph {et~al.}(2014)\citenamefont {Mazin},
  \citenamefont {Jeschke}, \citenamefont {Lechermann}, \citenamefont {Lee},
  \citenamefont {Fink}, \citenamefont {Thomale},\ and\ \citenamefont
  {Valent{\'{i}}}}]{Mazin2014}%
  \BibitemOpen
  \bibfield  {author} {\bibinfo {author} {\bibfnamefont {I.~I.}\ \bibnamefont
  {Mazin}}, \bibinfo {author} {\bibfnamefont {H.~O.}\ \bibnamefont {Jeschke}},
  \bibinfo {author} {\bibfnamefont {F.}~\bibnamefont {Lechermann}}, \bibinfo
  {author} {\bibfnamefont {H.}~\bibnamefont {Lee}}, \bibinfo {author}
  {\bibfnamefont {M.}~\bibnamefont {Fink}}, \bibinfo {author} {\bibfnamefont
  {R.}~\bibnamefont {Thomale}}, \ and\ \bibinfo {author} {\bibfnamefont
  {R.}~\bibnamefont {Valent{\'{i}}}},\ }\href {\doibase 10.1038/ncomms5261}
  {\bibfield  {journal} {\bibinfo  {journal} {Nature communications}\ }\textbf
  {\bibinfo {volume} {5}},\ \bibinfo {pages} {4261} (\bibinfo {year}
  {2014})}\BibitemShut {NoStop}%
\bibitem [{\citenamefont {Lanat{\`{a}}}\ \emph {et~al.}(2015)\citenamefont
  {Lanat{\`{a}}}, \citenamefont {Yao}, \citenamefont {Wang}, \citenamefont
  {Ho},\ and\ \citenamefont {Kotliar}}]{Lanata2015}%
  \BibitemOpen
  \bibfield  {author} {\bibinfo {author} {\bibfnamefont {N.}~\bibnamefont
  {Lanat{\`{a}}}}, \bibinfo {author} {\bibfnamefont {Y.~X.}\ \bibnamefont
  {Yao}}, \bibinfo {author} {\bibfnamefont {C.~Z.}\ \bibnamefont {Wang}},
  \bibinfo {author} {\bibfnamefont {K.~M.}\ \bibnamefont {Ho}}, \ and\ \bibinfo
  {author} {\bibfnamefont {G.}~\bibnamefont {Kotliar}},\ }\href {\doibase
  10.1103/PhysRevX.5.011008} {\bibfield  {journal} {\bibinfo  {journal}
  {Physical Review X}\ }\textbf {\bibinfo {volume} {5}},\ \bibinfo {pages}
  {11008} (\bibinfo {year} {2015})}\BibitemShut {NoStop}%
\bibitem [{\citenamefont {Piefke}\ and\ \citenamefont
  {Lechermann}(2018)}]{Piefke2017}%
  \BibitemOpen
  \bibfield  {author} {\bibinfo {author} {\bibfnamefont {C.}~\bibnamefont
  {Piefke}}\ and\ \bibinfo {author} {\bibfnamefont {F.}~\bibnamefont
  {Lechermann}},\ }\href {\doibase 10.1103/PhysRevB.97.125154} {\bibfield
  {journal} {\bibinfo  {journal} {Phys. Rev. B}\ }\textbf {\bibinfo {volume}
  {97}},\ \bibinfo {pages} {125154} (\bibinfo {year} {2018})}\BibitemShut
  {NoStop}%
\bibitem [{\citenamefont {Behrmann}\ and\ \citenamefont
  {Lechermann}(2015)}]{Behrmann2015}%
  \BibitemOpen
  \bibfield  {author} {\bibinfo {author} {\bibfnamefont {M.}~\bibnamefont
  {Behrmann}}\ and\ \bibinfo {author} {\bibfnamefont {F.}~\bibnamefont
  {Lechermann}},\ }\href {\doibase 10.1103/PhysRevB.91.075110} {\bibfield
  {journal} {\bibinfo  {journal} {Physical Review B}\ }\textbf {\bibinfo
  {volume} {91}},\ \bibinfo {pages} {075110} (\bibinfo {year}
  {2015})}\BibitemShut {NoStop}%
\bibitem [{\citenamefont {Ayral}\ \emph {et~al.}(2017)\citenamefont {Ayral},
  \citenamefont {Lee},\ and\ \citenamefont {Kotliar}}]{Thomas_DMET_RISB}%
  \BibitemOpen
  \bibfield  {author} {\bibinfo {author} {\bibfnamefont {T.}~\bibnamefont
  {Ayral}}, \bibinfo {author} {\bibfnamefont {T.-H.}\ \bibnamefont {Lee}}, \
  and\ \bibinfo {author} {\bibfnamefont {G.}~\bibnamefont {Kotliar}},\ }\href
  {\doibase 10.1103/PhysRevB.96.235139} {\bibfield  {journal} {\bibinfo
  {journal} {Phys. Rev. B}\ }\textbf {\bibinfo {volume} {96}},\ \bibinfo
  {pages} {235139} (\bibinfo {year} {2017})}\BibitemShut {NoStop}%
\bibitem [{\citenamefont {Lanat\`a}\ \emph {et~al.}(2015)\citenamefont
  {Lanat\`a}, \citenamefont {Deng},\ and\ \citenamefont
  {Kotliar}}]{Lanata_GA_finite_T}%
  \BibitemOpen
  \bibfield  {author} {\bibinfo {author} {\bibfnamefont {N.}~\bibnamefont
  {Lanat\`a}}, \bibinfo {author} {\bibfnamefont {X.}~\bibnamefont {Deng}}, \
  and\ \bibinfo {author} {\bibfnamefont {G.}~\bibnamefont {Kotliar}},\ }\href
  {\doibase 10.1103/PhysRevB.92.081108} {\bibfield  {journal} {\bibinfo
  {journal} {Phys. Rev. B}\ }\textbf {\bibinfo {volume} {92}},\ \bibinfo
  {pages} {081108} (\bibinfo {year} {2015})}\BibitemShut {NoStop}%
\bibitem [{\citenamefont {Wang}\ \emph {et~al.}(2010)\citenamefont {Wang},
  \citenamefont {He}, \citenamefont {Wang}, \citenamefont {Wang}, \citenamefont
  {Wang},\ and\ \citenamefont {Zhang}}]{Wang_GA_finite_T}%
  \BibitemOpen
  \bibfield  {author} {\bibinfo {author} {\bibfnamefont {W.-S.}\ \bibnamefont
  {Wang}}, \bibinfo {author} {\bibfnamefont {X.-M.}\ \bibnamefont {He}},
  \bibinfo {author} {\bibfnamefont {D.}~\bibnamefont {Wang}}, \bibinfo {author}
  {\bibfnamefont {Q.-H.}\ \bibnamefont {Wang}}, \bibinfo {author}
  {\bibfnamefont {Z.~D.}\ \bibnamefont {Wang}}, \ and\ \bibinfo {author}
  {\bibfnamefont {F.~C.}\ \bibnamefont {Zhang}},\ }\href {\doibase
  10.1103/PhysRevB.82.125105} {\bibfield  {journal} {\bibinfo  {journal} {Phys.
  Rev. B}\ }\textbf {\bibinfo {volume} {82}},\ \bibinfo {pages} {125105}
  (\bibinfo {year} {2010})}\BibitemShut {NoStop}%
\bibitem [{\citenamefont {Sandri}\ \emph {et~al.}(2013)\citenamefont {Sandri},
  \citenamefont {Capone},\ and\ \citenamefont {Fabrizio}}]{Sandri_finite_T}%
  \BibitemOpen
  \bibfield  {author} {\bibinfo {author} {\bibfnamefont {M.}~\bibnamefont
  {Sandri}}, \bibinfo {author} {\bibfnamefont {M.}~\bibnamefont {Capone}}, \
  and\ \bibinfo {author} {\bibfnamefont {M.}~\bibnamefont {Fabrizio}},\ }\href
  {\doibase 10.1103/PhysRevB.87.205108} {\bibfield  {journal} {\bibinfo
  {journal} {Phys. Rev. B}\ }\textbf {\bibinfo {volume} {87}},\ \bibinfo
  {pages} {205108} (\bibinfo {year} {2013})}\BibitemShut {NoStop}%
\bibitem [{\citenamefont {Sandhoefer}\ and\ \citenamefont
  {Chan}(2016)}]{DMET_e_phonon}%
  \BibitemOpen
  \bibfield  {author} {\bibinfo {author} {\bibfnamefont {B.}~\bibnamefont
  {Sandhoefer}}\ and\ \bibinfo {author} {\bibfnamefont {G.~K.-L.}\ \bibnamefont
  {Chan}},\ }\href {\doibase 10.1103/PhysRevB.94.085115} {\bibfield  {journal}
  {\bibinfo  {journal} {Phys. Rev. B}\ }\textbf {\bibinfo {volume} {94}},\
  \bibinfo {pages} {085115} (\bibinfo {year} {2016})}\BibitemShut {NoStop}%
\bibitem [{\citenamefont {Reinhard}\ \emph {et~al.}()\citenamefont {Reinhard},
  \citenamefont {Mordovina}, \citenamefont {Hubig}, \citenamefont {Kretchmer},
  \citenamefont {Schollwöck}, \citenamefont {Appel},\ and\ \citenamefont
  {Sentef}}]{DMET_e-p_Reinhard}%
  \BibitemOpen
  \bibfield  {author} {\bibinfo {author} {\bibfnamefont {T.~E.}\ \bibnamefont
  {Reinhard}}, \bibinfo {author} {\bibfnamefont {U.}~\bibnamefont {Mordovina}},
  \bibinfo {author} {\bibfnamefont {C.}~\bibnamefont {Hubig}}, \bibinfo
  {author} {\bibfnamefont {J.~S.}\ \bibnamefont {Kretchmer}}, \bibinfo {author}
  {\bibfnamefont {U.}~\bibnamefont {Schollwöck}}, \bibinfo {author}
  {\bibfnamefont {H.}~\bibnamefont {Appel}}, \ and\ \bibinfo {author}
  {\bibfnamefont {A.~A.}\ \bibnamefont {Sentef}, \bibfnamefont
  {Michael~Rubio}},\ }\href@noop {} {\bibinfo  {journal} {arXiv:1811.00048}\
  }\BibitemShut {NoStop}%
\bibitem [{\citenamefont {Lanat\`a}\ \emph {et~al.}(2017)\citenamefont
  {Lanat\`a}, \citenamefont {Lee}, \citenamefont {Yao},\ and\ \citenamefont
  {Dobrosavljevi\ifmmode~\acute{c}\else \'{c}\fi{}}}]{ghost_GA}%
  \BibitemOpen
\bibfield  {journal} {  }\bibfield  {author} {\bibinfo {author} {\bibfnamefont
  {N.}~\bibnamefont {Lanat\`a}}, \bibinfo {author} {\bibfnamefont {T.-H.}\
  \bibnamefont {Lee}}, \bibinfo {author} {\bibfnamefont {Y.-X.}\ \bibnamefont
  {Yao}}, \ and\ \bibinfo {author} {\bibfnamefont {V.}~\bibnamefont
  {Dobrosavljevi\ifmmode~\acute{c}\else \'{c}\fi{}}},\ }\href {\doibase
  10.1103/PhysRevB.96.195126} {\bibfield  {journal} {\bibinfo  {journal} {Phys.
  Rev. B}\ }\textbf {\bibinfo {volume} {96}},\ \bibinfo {pages} {195126}
  (\bibinfo {year} {2017})}\BibitemShut {NoStop}%
\bibitem [{\citenamefont {Fabrizio}(2007)}]{Fabrizio2007}%
  \BibitemOpen
  \bibfield  {author} {\bibinfo {author} {\bibfnamefont {M.}~\bibnamefont
  {Fabrizio}},\ }\href {\doibase 10.1103/PhysRevB.76.165110} {\bibfield
  {journal} {\bibinfo  {journal} {Phys. Rev. B}\ }\textbf {\bibinfo {volume}
  {76}},\ \bibinfo {pages} {165110} (\bibinfo {year} {2007})}\BibitemShut
  {NoStop}%
\bibitem [{\citenamefont {Zheng}()}]{Zheng_thesis}%
  \BibitemOpen
  \bibfield  {author} {\bibinfo {author} {\bibfnamefont {B.-X.}\ \bibnamefont
  {Zheng}},\ }\href {https://arxiv.org/abs/1803.10259} {\bibinfo  {journal}
  {arXiv:1803.10259}\ }\BibitemShut {NoStop}%
\bibitem [{\citenamefont {Pulay}(1980)}]{DIIS}%
  \BibitemOpen
\bibfield  {journal} {  }\bibfield  {author} {\bibinfo {author} {\bibfnamefont
  {P.}~\bibnamefont {Pulay}},\ }\href {\doibase
  https://doi.org/10.1016/0009-2614(80)80396-4} {\bibfield  {journal} {\bibinfo
   {journal} {Chemical Physics Letters}\ }\textbf {\bibinfo {volume} {73}},\
  \bibinfo {pages} {393 } (\bibinfo {year} {1980})}\BibitemShut {NoStop}%
\bibitem [{\citenamefont {Lieb}\ and\ \citenamefont {Wu}(1968)}]{Bethe}%
  \BibitemOpen
  \bibfield  {author} {\bibinfo {author} {\bibfnamefont {E.~H.}\ \bibnamefont
  {Lieb}}\ and\ \bibinfo {author} {\bibfnamefont {F.~Y.}\ \bibnamefont {Wu}},\
  }\href {\doibase 10.1103/PhysRevLett.20.1445} {\bibfield  {journal} {\bibinfo
   {journal} {Phys. Rev. Lett.}\ }\textbf {\bibinfo {volume} {20}},\ \bibinfo
  {pages} {1445} (\bibinfo {year} {1968})}\BibitemShut {NoStop}%
\bibitem [{\citenamefont {Capone}\ \emph {et~al.}(2004)\citenamefont {Capone},
  \citenamefont {Civelli}, \citenamefont {Kancharla}, \citenamefont
  {Castellani},\ and\ \citenamefont {Kotliar}}]{capone2004}%
  \BibitemOpen
  \bibfield  {author} {\bibinfo {author} {\bibfnamefont {M.}~\bibnamefont
  {Capone}}, \bibinfo {author} {\bibfnamefont {M.}~\bibnamefont {Civelli}},
  \bibinfo {author} {\bibfnamefont {S.~S.}\ \bibnamefont {Kancharla}}, \bibinfo
  {author} {\bibfnamefont {C.}~\bibnamefont {Castellani}}, \ and\ \bibinfo
  {author} {\bibfnamefont {G.}~\bibnamefont {Kotliar}},\ }\href {\doibase
  10.1103/PhysRevB.69.195105} {\bibfield  {journal} {\bibinfo  {journal} {Phys.
  Rev. B}\ }\textbf {\bibinfo {volume} {69}},\ \bibinfo {pages} {195105}
  (\bibinfo {year} {2004})}\BibitemShut {NoStop}%
\bibitem [{\citenamefont {Brinkman}\ and\ \citenamefont
  {Rice}(1970)}]{Brinkmann_Rice_1970}%
  \BibitemOpen
  \bibfield  {author} {\bibinfo {author} {\bibfnamefont {W.~F.}\ \bibnamefont
  {Brinkman}}\ and\ \bibinfo {author} {\bibfnamefont {T.~M.}\ \bibnamefont
  {Rice}},\ }\href {\doibase 10.1103/PhysRevB.2.4302} {\bibfield  {journal}
  {\bibinfo  {journal} {Phys. Rev. B}\ }\textbf {\bibinfo {volume} {2}},\
  \bibinfo {pages} {4302} (\bibinfo {year} {1970})}\BibitemShut {NoStop}%
\bibitem [{\citenamefont {Park}\ \emph {et~al.}(2008)\citenamefont {Park},
  \citenamefont {Haule},\ and\ \citenamefont {Kotliar}}]{Park_CDMFT}%
  \BibitemOpen
  \bibfield  {author} {\bibinfo {author} {\bibfnamefont {H.}~\bibnamefont
  {Park}}, \bibinfo {author} {\bibfnamefont {K.}~\bibnamefont {Haule}}, \ and\
  \bibinfo {author} {\bibfnamefont {G.}~\bibnamefont {Kotliar}},\ }\href
  {\doibase 10.1103/PhysRevLett.101.186403} {\bibfield  {journal} {\bibinfo
  {journal} {Phys. Rev. Lett.}\ }\textbf {\bibinfo {volume} {101}},\ \bibinfo
  {pages} {186403} (\bibinfo {year} {2008})}\BibitemShut {NoStop}%
\bibitem [{\citenamefont {Schir{\'{o}}}\ and\ \citenamefont
  {Fabrizio}(2010)}]{Schiro2010}%
  \BibitemOpen
  \bibfield  {author} {\bibinfo {author} {\bibfnamefont {M.}~\bibnamefont
  {Schir{\'{o}}}}\ and\ \bibinfo {author} {\bibfnamefont {M.}~\bibnamefont
  {Fabrizio}},\ }\href {\doibase 10.1103/PhysRevLett.105.076401} {\bibfield
  {journal} {\bibinfo  {journal} {Physical Review Letters}\ }\textbf {\bibinfo
  {volume} {105}},\ \bibinfo {pages} {076401} (\bibinfo {year}
  {2010})}\BibitemShut {NoStop}%
\bibitem [{\citenamefont {Schir{\'{o}}}\ and\ \citenamefont
  {Fabrizio}(2011)}]{Schiro2011}%
  \BibitemOpen
  \bibfield  {author} {\bibinfo {author} {\bibfnamefont {M.}~\bibnamefont
  {Schir{\'{o}}}}\ and\ \bibinfo {author} {\bibfnamefont {M.}~\bibnamefont
  {Fabrizio}},\ }\href {\doibase 10.1103/PhysRevB.83.165105} {\bibfield
  {journal} {\bibinfo  {journal} {Physical Review B}\ }\textbf {\bibinfo
  {volume} {83}},\ \bibinfo {pages} {165105} (\bibinfo {year}
  {2011})}\BibitemShut {NoStop}%
\bibitem [{\citenamefont {Mazza}\ and\ \citenamefont
  {Georges}(2017)}]{Mazza2017}%
  \BibitemOpen
  \bibfield  {author} {\bibinfo {author} {\bibfnamefont {G.}~\bibnamefont
  {Mazza}}\ and\ \bibinfo {author} {\bibfnamefont {A.}~\bibnamefont
  {Georges}},\ }\href {\doibase 10.1103/PhysRevB.96.064515} {\bibfield
  {journal} {\bibinfo  {journal} {Physical Review B}\ }\textbf {\bibinfo
  {volume} {96}},\ \bibinfo {pages} {064515} (\bibinfo {year}
  {2017})}\BibitemShut {NoStop}%
\bibitem [{\citenamefont {Kretchmer}\ and\ \citenamefont
  {Chan}(2018)}]{Kretchmer2018}%
  \BibitemOpen
  \bibfield  {author} {\bibinfo {author} {\bibfnamefont {J.~S.}\ \bibnamefont
  {Kretchmer}}\ and\ \bibinfo {author} {\bibfnamefont {G.~K.-L.}\ \bibnamefont
  {Chan}},\ }\href {\doibase 10.1063/1.5012766} {\bibfield  {journal} {\bibinfo
   {journal} {The Journal of Chemical Physics}\ }\textbf {\bibinfo {volume}
  {148}},\ \bibinfo {pages} {054108} (\bibinfo {year} {2018})},\ \Eprint
  {http://arxiv.org/abs/https://doi.org/10.1063/1.5012766}
  {https://doi.org/10.1063/1.5012766} \BibitemShut {NoStop}%
\bibitem [{\citenamefont {Fertitta}\ and\ \citenamefont
  {Booth}(2018)}]{Booth2018}%
  \BibitemOpen
  \bibfield  {author} {\bibinfo {author} {\bibfnamefont {E.}~\bibnamefont
  {Fertitta}}\ and\ \bibinfo {author} {\bibfnamefont {G.~H.}\ \bibnamefont
  {Booth}},\ }\href {\doibase 10.1103/PhysRevB.98.235132} {\bibfield  {journal}
  {\bibinfo  {journal} {Phys. Rev. B}\ }\textbf {\bibinfo {volume} {98}},\
  \bibinfo {pages} {235132} (\bibinfo {year} {2018})}\BibitemShut {NoStop}%
\end{thebibliography}%

\end{document}